\documentclass[aps, prb, amsfonts, amsmath, twocolumn, superscriptaddress, 10pt, floatfix]{revtex4-1}
\usepackage{natbib}

\usepackage[utf8]{inputenc}
\usepackage[english]{babel}
\usepackage{amsmath}
\usepackage{amsfonts}
\usepackage{amssymb}
\usepackage{graphicx}
\usepackage{xcolor}

\begin{document}

%\title{Role of the excited manifold in the energy transfer in a quantum circuit simulator of a three-site light-harvesting complex.}
\title{The role of the multiple excitation manifold in a driven quantum simulator of an antenna complex}

\author{A. W. Chin}
\affiliation{Institut des NanoSciences de Paris, Sorbonne Universit\'e, 4 place Jussieu, boite courrier 840, 75252 PARIS Cedex 05, France}

\author{B. Le Dé}
\affiliation{Universit\'e Paris Saclay, CNRS, Institut de Chimie Physique UMR8000, 91405, Orsay, France}

\author{E. Mangaud}
\affiliation{Institut des NanoSciences de Paris, Sorbonne Universit\'e, 4 place Jussieu, boite courrier 840, 75252 PARIS Cedex 05, France}
\affiliation{Laboratoire Collisions Agr\'egats R\'eactivi\'e (IRSAMC), Universit\'e Toulouse III Paul Sabatier, UMR 5589, F-31062 Toulouse Cedex 09, France}	

\author{O. Atabek}
\affiliation{Universit\'e Paris Saclay, CNRS, Institut des Sciences Mol\'eculaires d'Orsay, 91405, Orsay, France}

\author{M. Desouter-Lecomte}
\affiliation{Universit\'e Paris Saclay, CNRS, Institut de Chimie Physique UMR8000, 91405, Orsay, France}
\affiliation{D\'epartement de Chimie, Universit\'e de Li\`ege, Sart Tilman, B6, B-4000 Li\`ege, Belgium}

\begin{abstract}

Biomolecular light-harvesting antennas operate as nanoscale devices in a regime where the coherent interactions of individual light, matter and vibrational quanta are non-perturbatively strong. The complex behaviour arising from this could, if fully understood, be exploited for myriad energy applications. However, non-perturbative dynamics are computationally challenging to simulate, and experiments on biomaterials explore very limited regions of the non-perturbative parameter space. So-called `quantum simulators' of light-harvesting models could provide a solution to this problem, and here we employ the hierarchical equations of motion technique to investigate recent superconducting experiments  of Poto{\v{c}}nik $\it{et}$ $\it{al.}$ (Nat. Com. 9, 904 (2018)) used to explore  excitonic energy capture. By explicitly including the role of optical driving fields, non-perturbative dephasing noise and the full multi-excitation Hilbert space of a three-qubit quantum circuit, we predict the measureable impact of these factors on transfer efficiency. By analysis of the eigenspectrum of the network, we uncover a structure of energy levels that allows the network to exploit optical `dark' states and excited state absorption for energy transfer. We also confirm that time-resolvable coherent oscillations could be experimentally observed, even under strong, non-additive action of the driving and optical fields.    

 \end{abstract}

\maketitle

\section{INTRODUCTION}
Photosynthetic pigment-protein complexes (PPCs) are bioengineered optoelectronic 'devices' that perform crucial light-harvesting tasks such as spatially directed excitonic energy transport (EET) and highly efficient exciton-to-charge generation \cite{blankenship2014molecular}. Understanding how this is achieved in these self-assembling nano-systems could lead to fundamentally new approaches for sustainable photovoltaic and catalytic technologies, and interest in this topic has been further stoked by the intriguing but highly controversial proposal that quantum coherence and entanglement might play a role in these biological functions \cite{Scholes2017,chin2013role,Collini2010,fuller2014vibronic,Kreisbeck2012,Lambert_2013,Lee1462,Panitchayangkoon2010,romero2014quantum, duan2017nature,maiuri2018coherent}. For instance, some recent work showed that coherence could not be a feature selected for during evolution \cite{valleau2017}. While debate over this latter aspect is ongoing, the essential idea of engineering advantageous quantum dynamics into functional molecular materials is presently being developed in a range of less complex, man-made organic systems, such as those found in molecular photovoltaics  \cite{gelinas2014ultrafast, smith2015phonon,bredas2017photovoltaic}, polaritons \cite{feist2017polaritonic,del2018tensor} and a range of theoretical proposals for photocells based on quantum heat engines \cite{scully2011quantum,creatore2013efficient,wertnik2018optimizing}.      
 
However, observing and elucidating the mechanisms of ultrafast (fs-ps) EET in organic materials is a very challenging experimental task, and often requires advanced nonlinear optical experiments that can only be performed on inhomogeneous ensembles of nanostructures. Experiments capable of addressing single PPCs or nanostructures have been demonstrated, but are typically restricted to certain probes, such as fluorescence \cite{hildner2011femtosecond,hildner2013quantum}. At the same time, theoretical studies also point to a very wide range of electronic and environmental factors that can contribute to rapid and directed EET, among which the roles of non-perturbative and non-Markovian vibrational dephasing noise, and electronic disorder are particularly important \cite{Chin2013,novelli2015vibronic,lim2015vibronic,Schulze2015explicit,Valkunas_vib2016,Chen2016,oviedo2016phase,Troisi2017,Grondelle2017,csurgay2018}.  In order to take these latter features into account requires advanced and numerically expensive computational techniques for simulating open quantum dynamics \cite{Ishizaki_2009,Akihito09,Castro2014,Chen2015,Chin2012,Chin2013,Dijkstra_2010,elinor1,Iles_Smith_2015,killoran2015,Lee1462,Mal__2016,Qin2017,Santamore2013,Stones2016,schroder2019tensor}, and including the full quantum mechanics of the PPC light-matter interaction adds enormously to this problem. This results in exponentially scaling demands on computing resources as the number of pigments increases. Numerical simulations are still feasible for proteins such as the seven-pigment Fenna-Matthews-Olson complex. However, a light-harvesting antenna such as a chlorosome with it's $10^5$ quantum two-level systems requires very efficient computing strategies and capabilities \cite{fujita2014,sawaya2015,huh2014}. Building and simulating atomistically realistic quantum models of exciton transport in PPCs can therefore also be as challenging as experimentation, and in some cases even more difficult.

A potentially powerful solution to these problems is offered by the emergence of so-called 'quantum simulators' \cite{Mostame2012,Mostame2016}. These systems allow a complex quantum system to be 'simulated' by building an analogue of their underlying microscopic models with controllable quantum bits (qubits). Performing experiments on these tunable and rationally designed platforms permits access to the key physics of the system across a large parameter space and - when scaled up to include many qubits - could allow for simulations of models that would be impossible on classical computers, as has been recently demonstrated with superconducting circuits \cite{arute2019quantum}. Recently, a prototype of an experimental quantum simulator for molecular exciton transport has been demonstrated by Poto{\v{c}}nik $\it{et}$ $\it{al.}$ \cite{potovcnik2018studying}. This system is based on a coupled three qubit-chromophore setup similar to the more general superconducting networks proposed theoretically by Mostame et al. in Refs. \cite{Mostame2012,Mostame2016}. Other light-harvesting simulators using either nuclear spins or ion traps have also recently appeared \cite{bi2018efficient, gorman2018engineering}.  Here, we shall focus on parameters relevant to the superconducting experiments in Ref. \cite{potovcnik2018studying}. This setup enables the impact of many independent degrees of freedom to be examined, including 'system properties' such as the inter-site coherent coupling and local energy gaps, as well as 'environmental' properties, such as the dissipative noise coupling and spectral density. A clever scheme for selectively exciting the qubits with wave-guided microwaves (MW) and a site-selective extraction of transported excitation energy through a MW resonator also allow this artificial 'PPC' to be explored in highly non-natural conditions, such as strong light-matter coupling, single photon excitation and non-classical photonic pumping. We also note that this few-qubit systems with tunable \emph{environmental} parameters could also be a promising platform for exploring ideas related to quantum thermodynamics. 

In a recent work, we have used numerically exact hierarchical equations of motion (HEOM)\cite{Tanimura_1989,Tanimura_2006,Ishizaki_2005,Ishizaki_2009,Xu_2007,Shi_2009,Shi2012,Shi_2014,Schulten_2012,Kreisbeck2014} to predict the ultrafast generation of electronic coherences during the incoherent relaxation of a high-lying excitonic state into a doublet of nearly degenerate low-lying states \cite{Chin_qubit2018}. This novel process could be realized in the particular configuration of the SC qubit circuit of Poto{\v{c}}nik $\it{et}$ $\it{al.}$\cite{potovcnik2018studying} in the following way: Inter site couplings and gaps are chosen so that the first excited manifold consists in a bright state (optically excitable) separated from a doublet of dark (nonradiative) states such that the bright-dark energy gap is in resonance with a sharply peaked spectral density of the noise. This situation, known as vibrationally assisted electronic decay or 'phonon antenna' due to a sharply structured spectral density, has been studied in the context of EET in several different contexts \cite{rey2013exploiting, irish2014vibration, Chin2013,kolli2012fundamental,Chin2019}.  The bath induced population-to-coherence process already predicted by nonsecular Redfield theory has been confirmed by HEOM simulations in the strong coupling regime \cite{Chin_qubit2018}. We showed that this generation of electronic coherence is the most efficient for a regime between weak and strong system-bath couplings, which is a general observation across the various types of biological 'noise-assisted ' transport that have been studied. The generated quantum superposition of the dark doublet corresponds to an oscillatory energy transport across the sites in 'real space' that could be detected by the resonator emission. This behavior also survives when the noise is classically stochastic, i.e. when the real part of the correlation function of the bath mode dominates the imaginary part so that the classical, rather than quantum, fluctuation-dissipation relation is satisfied (\textit{vide infra}).  

\begin{figure*}
\centering

\includegraphics[width =1.\textwidth]{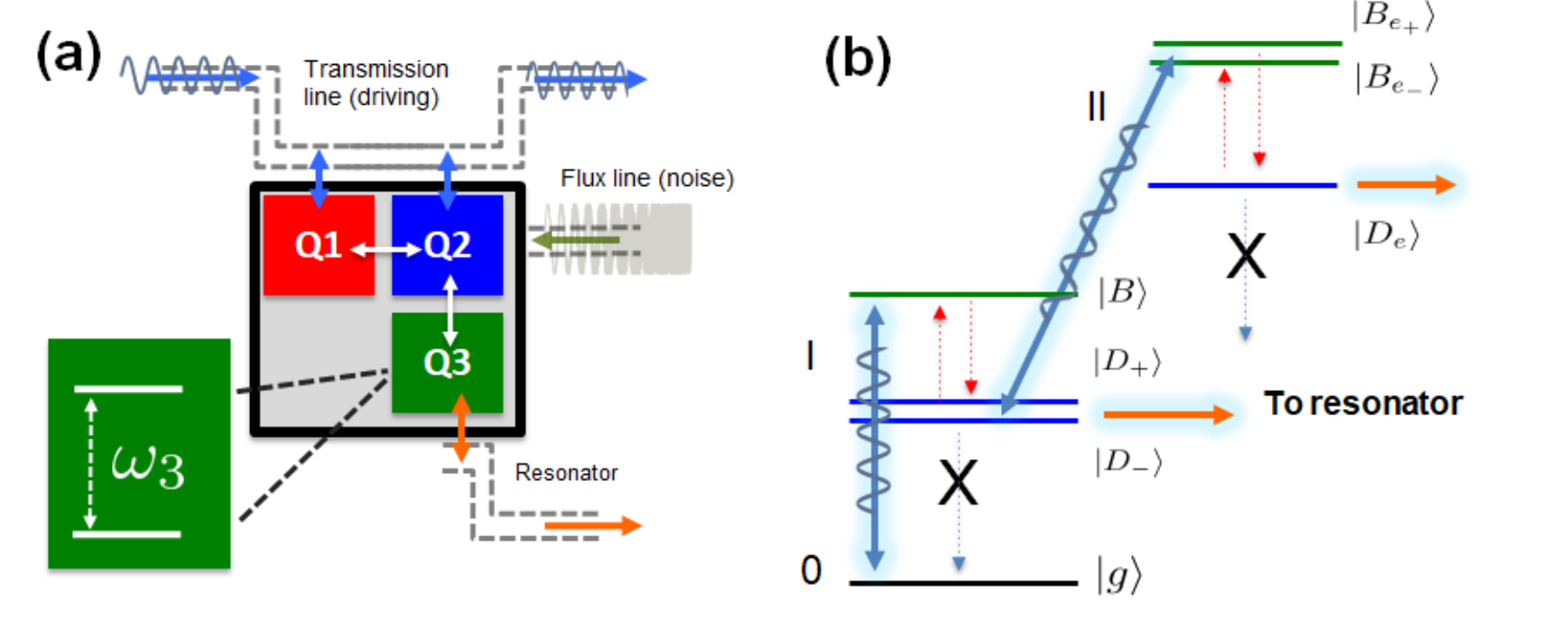}
\caption{ (a) Schematic representation of the superconducting quantum circuit used in Ref. \cite{potovcnik2018studying} to simulate energy transport in a photosynthetic light-harvesting array. Here, three qubits ($Q_1, Q_2, Q_3$) act as chromophores with a tunable excitation energy $\omega_{i}$ and are coupled together by nearest-neighbor capacitive interactions (white arrows). Qubits $Q_1$ and $Q_2$ are coupled identically to a transmission line (blue arrows) which carries the excitation/pump fields, while emission in the resonator line is only sensitive to the excitation of qubit $Q_3$. The flux lines are used to tune $\omega_{i}$, allowing the application of stochastic signals to mimic an arbitrary classical dephasing noise on the qubit (chromophore) system. (b) The resulting eigenspectrum showing the one-excitation bright $|B\rangle$ and dark $|D_{\pm}\rangle$ eigenstates and of the higher-lying (two-excitation) excited manifold, consisting of the dark $|D_e\rangle$ and bright $|B_{e\pm}\rangle$ states. Due to interefence effects, the dark states cannot radiate into the transmission line and so only decay into the resonator. Dissipative transitions induced by noise (dashed red arrows) rapidly populate these states from the one and two-excitation bright states, leading to a ratchet-like transfer of energy from waveguide to resonator. The eigenspectrum of the system ensures that the dipole-allowed optical transitions (blue wavy arrows) are both resonant with the MW pulse in the waveguide, allowing effective pumping into the two-excitation sector. }
\label{fig1}
\end{figure*}

However, this previous work only looked at the single excitation dynamics on the qubit network, whereas a number of recent works have pointed to potentially richer quantum effects - such as superabsorption -  when the multi-excitation states are included \cite{higgins2014superabsorption}.  In this work, we return to the problem with a significantly improved description of the entire three-qubit (8-level) system, allowing us to explore much more of the experimental parameter space available than was considered in Ref. \cite{Chin_qubit2018}, and also to target effects arising from the `non-additivity' of the competing photonic and dephasing environments \cite{schroder2019tensor,wertnik2018optimizing, maguire2019environmental}. To be able to describe such physics requires a numerically exact treatment of the driven system and its dissipative environment, and we shall again make use of the non-perturbative HEOM method to account for the latter.  Importantly, we now also include spontaneous optical decay, temporally shaped driving fields and the full manifold of higher lying states, allowing the correlated multi-quanta dynamics of this `dark state photocell'  configuration of qubits to be analysed and optimised \cite{creatore2013efficient}. Using these new capabilities, we show that the Hamiltonian structure of the system in Ref.  \cite{Chin_qubit2018} permits a ratchet-like type of dark state protection to operate in the multi-excitation sectors of the model, as shown in Fig.\ref{fig1}. However, the dissipative trapping of excitations in `dark' states does not increase the transport efficiency but only the absolute rates of energy capture. We shall also show that the maximum efficiency appears in the limit of low-temperature `quantum' noise and discuss the role of different driving protocols on the measurable energy transfer efficiencies of this light-harvesting simulator.

The paper is organized as follows. Section II describes the three-site simulator and the system-bath interaction in classical and quantum regimes.  Section III presents our numerical results for a classical or quantum noise and Section IV provides some discussions and perspectives for future investigations.A summary of the operational HEOM equations for system bath dynamics and of Lindblad terms \cite{Breuer2002} accounting for spontaneous emission is given in Appendix \ref{appendix:METH}.

\section{Model and parameters}
\subsection{The system-bath Hamiltonian}
The simulator of the excitonic Hamiltonian interacting with both a radiation and a dephasing environment is schematized in Fig.1. It contains three SC qubits simulating two-level systems (describing the ground and first excited local electronic states of three chromophore sites) with energy gaps $\hbar {\omega }_{i}$. This is exactly the same qubit geometry implemented by Poto{\v{c}}nik $\it{et}$ $\it{al.}$ \cite{potovcnik2018studying}, and so we shall use the same qubit parameters and the same ranges of noise and driving strengths as in the experimental setup. In particular, the setup allows the realisation of a sharply peaked (Lorentzian) spectral density or of a broadband white noise (see Fig.2 of Ref. \cite{potovcnik2018studying}). We focus here on the sharply structured case where the peak is in resonance with a main transition of the system. 
The three qubit frequencies are ${{\omega }_{1}}={{\omega }_{2}}=$12GHz, ${{\omega }_{3}}=$11.5GHz. Two qubits ${{Q}_{1}}$ and ${{Q}_{2}}$ are spatially close and coupled with the same strength to the transmission line while the third qubit ${{Q}_{3}}$ is separated from the two others and not directly excited by the electromagnetic field. It is linked to the resonator which collects the flux transmitted by the excitation transfer. The resonator emission is experimentally used to quantify the efficiency of the EET process through the network. The inter-site coherent coupling ${{J}_{ij}}$ is very strong between ${{Q}_{1}}$ and ${{Q}_{2}}$ while it is weak between ${{Q}_{2}}$ and ${{Q}_{3}}$ and negligible between ${{Q}_{1}}$ and ${{Q}_{3}}$ (${{J}_{12}}$ = 500MHz, ${{J}_{23}}=$ 50MHz and ${{J}_{13}}$ = 0MHz). The `noise' consists of fluctuations in the energy gaps of the qubits and, as is common for open quantum systems, is considered to arise from a bosonic bath of harmonic oscillators. For simplicity, and following the experiment of Ref. \cite{potovcnik2018studying}, the qubit noise is coupled to the excited state of qubit ${{Q}_{2}}$ only. By varying the effective temperature of the environment (\textit{vide infra}), we may simulate a classical or quantum bath.

The Hamiltonian in atomic units (with $\hbar = 1$) in the basis set of the qubit states (local site basis set, in terms of the Pauli $\sigma$ matrices) reads: $H={{H}_{S}}+{{H}_{f}}+{{H}_{ren}}+{{H}_{SB_a}}+{{H}_{B_a}}$ with
\begin{equation}
		{{H}_{S}}=\sum\nolimits_{i=1}^{3}{\frac{{{\omega }_{i}}}{2}}\sigma _{z}^{(i)}+\sum\limits_{i<j}{{{J}_{ij}}}\left( \sigma _{+}^{(i)}\sigma _{-}^{(j)}+\sigma _{+}^{(j)}\sigma _{\_}^{(i)} \right)
\end{equation}
and ${{H}_{field}}=-\sum\nolimits_{i=1}^{2}{\mu E(t)\sigma _{x}^{(i)}}$. The dipole operator induces transition between the two states of qubits ${{Q}_{1}}$ and ${{Q}_{2}}$. As only the Rabi frequency $\Omega (t)=\mu E(t)$ is important, $\mu $ is taken equal to 1 a.u. in the simulations. 

The bath is a collection of harmonic oscillators ${{H}_{B_a}}=\sum\limits_{k}{{{\omega }_{k}}}a_{k}^{\dagger }{{a}_{k}}$ written as a function of the bosonic creation and annihilation operators. It is linearly coupled to the system: ${{H}_{SB_a}}=S.B_a$ where $B_a=\frac{1}{\sqrt{2}}\sum\limits_{k}{{{g}_{k}}\left( {{a}_{k}}+a_{k}^{\dagger } \right)}$ is the collective bath mode and $S=\left( \sigma _{z}^{(2)}+\mathbf{1} \right)$ is the system coupling operator. $S$ specifies that only the excited state of qubit ${{Q}_{2}}$ is coupled to the noise that makes the energy gap fluctuate. ${H}_{ren}$ is a renormalization term due to the system-bath coupling inducing an energy shift ${H}_{ren}=\lambda =1/2\sum\nolimits_{k}{g_{k}^{2}}/{{\omega }_{k}}$.

Without a driving field and without coupling to the noise, the eigenstates of ${{H}_{S}}$ form two important excited triplets. The lower lying cluster contains a bright state $\left| B \right\rangle $ which can be excited by the transmission line and a doublet of dark states $\left| {{D}_{\pm }} \right\rangle $. The corresponding eigenvectors are given in Appendix \ref{appendix:EV}. Their expressions show that the dipoles of the sites interfere constructively for state $\left| B \right\rangle $ leading to the bright character. They are in opposition in the dark states which are decoupled from the field.  In the higher lying excited triplet, the characteristics of the states are inverted leading to a bright doublet $\left| B_{e\pm} \right\rangle $ and a single dark state $\left| {{D}_{e}} \right\rangle $. As seen in this scheme, the gap (12.5GHz) between the ground state and the bright state $\left| B \right\rangle $ is similar to that between the dark doublet $\left| D_{\pm} \right\rangle $ and the excited bright doublet $\left| {{B}_{e\pm }} \right\rangle $. The corresponding transition dipoles are ${{\mu }_{gB}}$ = 1.41 a.u., ${\mu }_{{D}_{-}B_{e-}}$= 0.74 a.u.,  ${\mu }_{{D}_{-}B_{e+}}$= 0.68 a.u., ${\mu }_{{D}_{+}B_{e-}}$= 0.73 a.u. and ${\mu }_{{D}_{+}B_{e+}}$= 0.66 a.u.(setting the individual, uncoupled qubits to each have transition dipole moments of  $\mu $ = 1 a.u.). This opens the possibility of populating the second excited doublet by a two-photon transition from the lower dark doublet populated via the bath since the carrier frequency used is in resonance with the $gB$ transition (12.5GHz). It is also possible to observe some transitions between the bright states since the transition dipole is favourable (${{\mu }_{B{{D}_{-}}}}$= 0.95 a.u. and ${{\mu }_{B{{D}_{+}}}}$= 1.04 a.u.) but the energy gap (11.5GHz) is no more in resonance with the carrier frequency. 

After diagonalization, all the eigenstates are coupled via the bath but the spectral density of the noise is sharply peaked at the $B{{D}_{\pm }}$ or $B_{e \pm}{{D}_{e}}$ transition (1GHz).  Excitation of the delocalized bright eigenstate initiates the transport by relaxing to the dark doublet. It is the heart of the setup as discussed in our previous work. The decay from the bright state may be considered as a quasi incoherent process but it creates population and coherence in the dark doublet at an equal rate. Due to the shape of the spectral density this doublet is protected from the environment. This superposed state can then be excited towards the higher lying brigth doublet.

\subsection{Classical versus quantum noise}
\begin{figure}
 \centering
\includegraphics[width =1.\columnwidth]{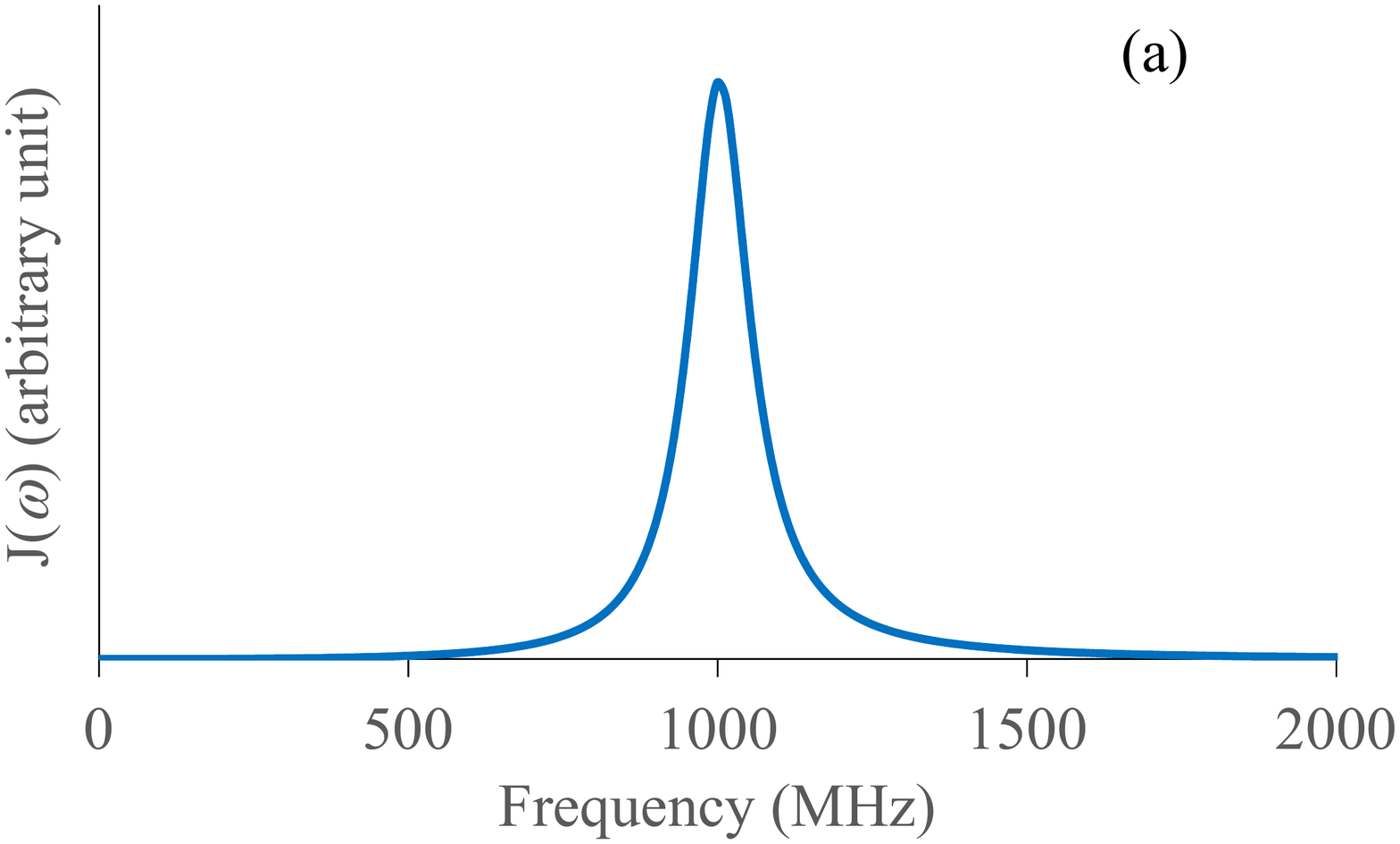}	
\includegraphics[width =1.\columnwidth]{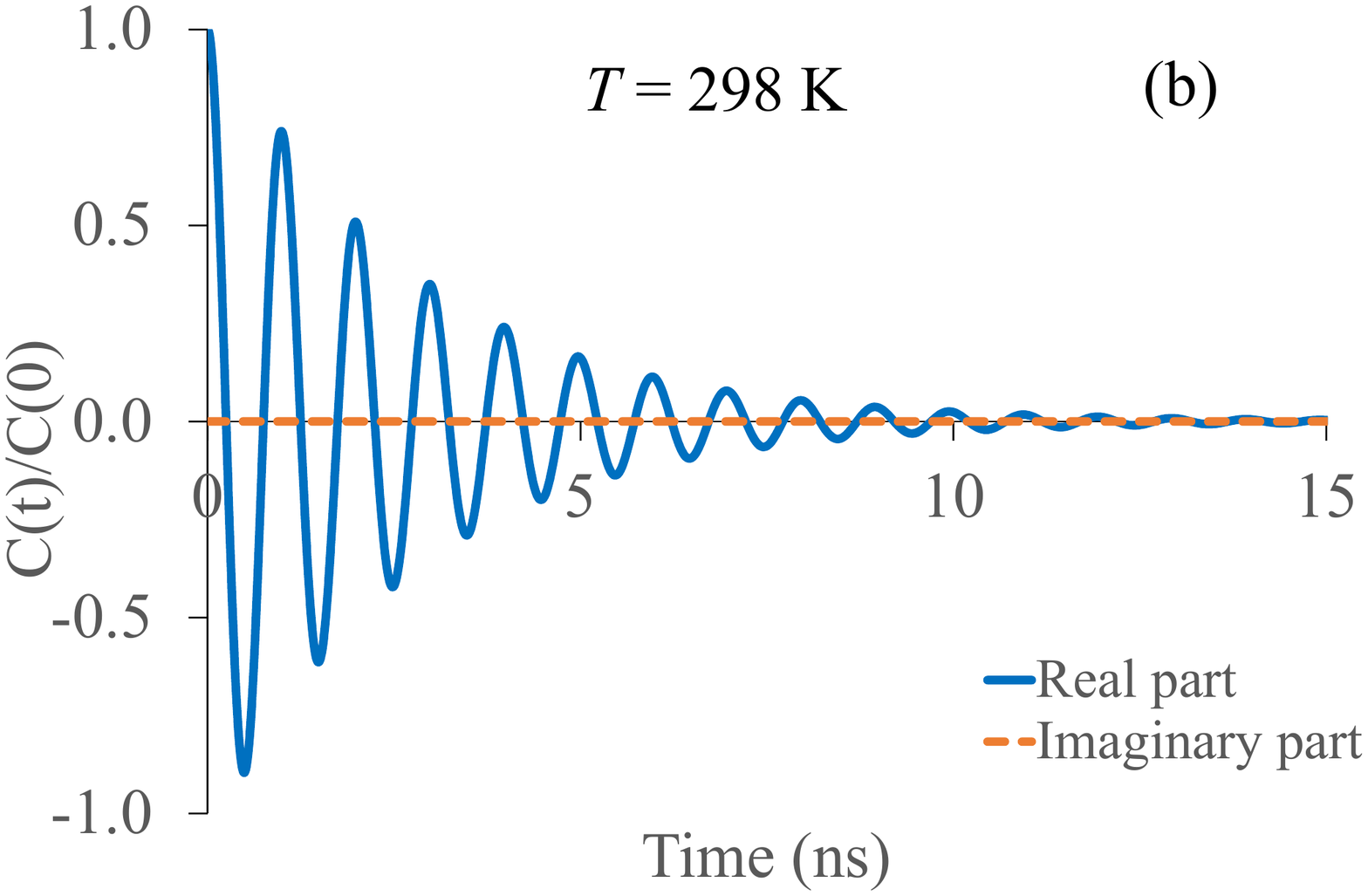}
\caption{Panel (a): Spectral density centered at the frequency $\omega_{BD_{\pm }}$. Panel (b): Normalized bath correlation function $C(t)$ (Eq.(\ref{ct}) with $\tau $ = 0) at room temperature corresponding to a classical noise in the present study. }
\label{fig:corre}
\end{figure}

In the simulator of Poto{\v{c}}nik $\it{et}$ $\it{al.}$, the noise is a classical stochastic signal allowing the generation of different power spectra (white or colored noise). However, we shall also discuss quantum noise by considering very low temperatures. The main tool in open quantum systems is the bath spectral density 
\begin{equation}
J(\omega )=\left( \pi /2 \right)\sum\nolimits_{k}{\left( g_{k}^{2}/{{\omega }_{k}} \right)}\delta (\omega -{{\omega }_{k}}).
\label{Jomega}
\end{equation}
The correlation function of the collective coordinate $B_a$ over the equilibrium bath at a given temperature is : 
\begin{equation}
		C\left( {t} \right) = \frac{1}{\pi }\int\limits_{ - \infty }^{ + \infty } {d\omega J\left( \omega  \right) n\left( \omega \right) e^{i\omega \left( t \right)}} .	
\label{ct}
\end{equation}
where $n(\omega )=1/\left( {{e}^{\beta \omega }}-1 \right)$ is the Bose function with $\beta =1/{{k}_{B}}T$ and ${{k}_{B}}$ is the Boltzmann constant. In the quantum regime, the correlation function is complex valued but when $\beta \to 0$ (for high enough temperature) the imaginary part becomes negligible with respect to the real part. In practice, due to the energy gaps considered here, room temperature is already the very high temperature limit, i.e., the noise is classical. In order to compare with a quantum noise, the temperature must be decreased to $T$= 0.01K. In the previous work  \cite{Chin_qubit2018}, we have used the dimensionless parameter 
\begin{equation}
\eta =\lambda /\hbar \omega_{BD_{\pm }}
\label{eta}
\end{equation}
to estimate the optimized coupling strength in quantum regime generating the superposition in the lower dark doublet. The maximum of efficiency was for  $\eta \approx $ 0.015. This parameter is less useful in the classical regime where the Bose function increases linearly with $T$ ($n(\omega )\to {{k}_{B}}T/\omega $). We rescale the coupling to the bath and therefore the $\eta $ parameter so that the decay rate estimated at the Golden rule approximation ${{R}_{{{D}_{\pm }}B}}=2\pi {{\left| {{V}_{{{D}_{\pm }}B}} \right|}^{2}}J(\hbar \omega_{BD_{\pm }})\left[ n(\hbar \omega_{BD_{\pm }})+1 \right]$ remains of the same order of magnitude in both regimes. $J(\omega )$ is multiplied by a factor $\hbar \omega_{BD_{\pm }}/{{k}_{B}}T$ to compensate for the diverging thermal populations in the high temperature limit. The peaked spectral density is displayed in Fig.\ref{fig:corre}a. It is fitted by a four-pole Lorentzian expression
\begin{equation}
J\left( \omega  \right)=\frac{p{{\omega }^{3}}}{{{\Lambda }_{1}}({{\Omega }_{1}},{{\Gamma }_{1}}){{\Lambda }_{2}}({{\Omega }_{2}},{{\Gamma }_{2}})}        
\label{J}
\end{equation}
where ${{\Lambda }_{k}}=\left[ {{\left( \omega +{{\Omega }_{k}} \right)}^{2}}+\Gamma _{k}^{2} \right]\left[ {{\left( \omega -{{\Omega }_{k}} \right)}^{2}}+\Gamma _{k}^{2} \right]. $
The numerical values of the parametrization are given in Appendix \ref{appendix:spectraldensity}. Fitting the spectral density Eq.(\ref{Jomega}) with the function given by Eq.(\ref{J}) leads to an analytical expression of $C(t)$:
\begin{equation}
C\left( {t} \right) = \sum\limits_{k = 1}^{{n_{cor}}} {{\alpha _k}{e^{i{\gamma _k}\left( {t} \right)}}} 
\label{Corre}
\end{equation}
Explicit expressions of the ${\alpha _k}$ and ${\gamma _k}$ can be found in the Appendix of Ref. \cite{Mangaud_2017}. ${n_{cor}}$ is the sum of the four terms coming from the four simple poles in the upper complex plane and, the terms related to the poles (Matsubara frequencies) of the Bose function. The complex conjugate of the correlation function can be expressed by keeping the same coefficients ${\gamma _k}$ in the exponential functions with modified coefficients ${\tilde \alpha _k}$ according to :
\begin{equation}
{C^*}\left( {t} \right) = \sum\limits_{k = 1}^{{n_{cor}}} {{{\tilde \alpha }_k}} {e^{i{\gamma _k}\left( {t} \right)}}
\label{Correconj}
\end{equation}
with	${\tilde \alpha _{1}} = \alpha _{2}^*$, ${\tilde \alpha _{2}} = \alpha _{1}^*$, ${\tilde \alpha _{3}} = \alpha _{4}^*$, ${\tilde \alpha _{4}} = \alpha _{3}^*$ where the indices $k = 1,4$ are related to the four poles of the superohmic Lorentzian function. The terms with $k > 4$ refer to the  Matsubara terms and then ${\tilde \alpha _{k}} = {\alpha _{k}}$   \cite{Tannor_2010}. Fig.\ref{fig:corre}b gives the real and imaginary parts of $C(t)/C(0)$ at room temperature. As expected from the sharp peak of $J(\omega)$ the correlation time is longer than a typical Rabi oscillation in one qubit. In the quantum regime, the imaginary part has the same order of magnitude than the real one and is simply out of phase. 

\section{FIELD DRIVEN DYNAMICS}
The dynamics of the reduced density matrix of the system after tracing on the bath degrees of freedom ${{\rho }_{S}}(t)=T{{r}_{B_a}}\left[ \rho_{tot} (t) \right]$ are treated by HEOM which can in principle take into account non-Markovian effects in a numerically exact manner. As the correlation time is here longer than the characteristic system timescale, some memory effects are expected. This point has already been discussed in our previous work where we have shown that a non-Markovian master equation is necessary to correctly describe the noise induced transfer of population to coherence \cite{Chin_qubit2018}. In this work, we want to better quantify the energy given to the resonator by the final site ${{Q}_{3}}$ and the radiative loss from the bright state. We therefore add to the HEOM master equation two Lindblad terms describing the spontaneous emission. The numerical methods are summarized in Appendix \ref{appendix:METH}.

We have compared dynamics with and without the coupling to the noise $H_{SB_a}$ while retaining the spontaneous emission process in order to emphasize the crucial role of the bath in assisting transfer of excitations towards ${{Q}_{3}}$.   The system-bath coupling is calibrated for the energy domain simulated here in order to match the optimal situation generating coherence in the lower dark doublet. In the quantum regime, we take a coupling so that $\eta=0.01$ (Eq.(\ref{eta})) and in the classical case, taking into account the  $\hbar \omega_{BD_{\pm }}/{{k}_{B}}T$ factor, $\eta=10^{-6}$. The emission rates are chosen equal to ${{G}_{3}}={{G}_{12}}$=10MHz (a reasonable value according to Ref.\cite{potovcnik2018studying}). In each simulation, the initial state is the ground state. The pulses have a simple sine square envelope and the carrier frequency correponds to the excitation of the lower bright state (12.5GHz)
\begin{equation}
E(t)=A {{\sin }^{2}}\left( \pi t/{{\tau }_{\max }} \right)\cos ({{\omega }_{gB}}t).
\label{E(t)j}
\end{equation}
For the simulations, a duration of about $\tau_{sim}$=500ns (longer than the pulse duration) has been considered, allowing the system to return to the ground state by spontaneous emission as described by the Lindblad terms ${{L}_{3}}({{\rho }_{S}}(t))$ and ${{L}_{12}}({{\rho }_{S}}(t))$. We compare different pulse energies: 
\begin{equation}
\mathcal{E}=\int_{0}^{{{\tau }_{\max }}}{{{E}^{2}}(t)dt} 
\label{I}
\end{equation}
and for each pulse duration ${{\tau }_{\max }}$ we adapt the corresponding maximum field amplitude $A$ such that the integrated intensity (Eq.\ref{I}) remains constant. We analyze the ratio
\begin{equation}
R=\frac{{{P}_{res}}}{{{P}_{res}}+{{P}_{loss}}} 
\label{R}
\end{equation}
where ${{P}_{res}}=\int_{0}^{{{\tau }_{sim }}}{{{L}_{3}}\left[ {{\rho }_{S}}(t) \right]dt}$ and ${{P}_{loss}}=\int_{0}^{{{\tau }_{sim }}}{{{L}_{12}}\left[ {{\rho }_{S}}(t) \right]dt}$.

\subsection{Classical noise}
\label{sec:cla_noise}

\begin{figure}
\centering
\includegraphics[width =1.\columnwidth]{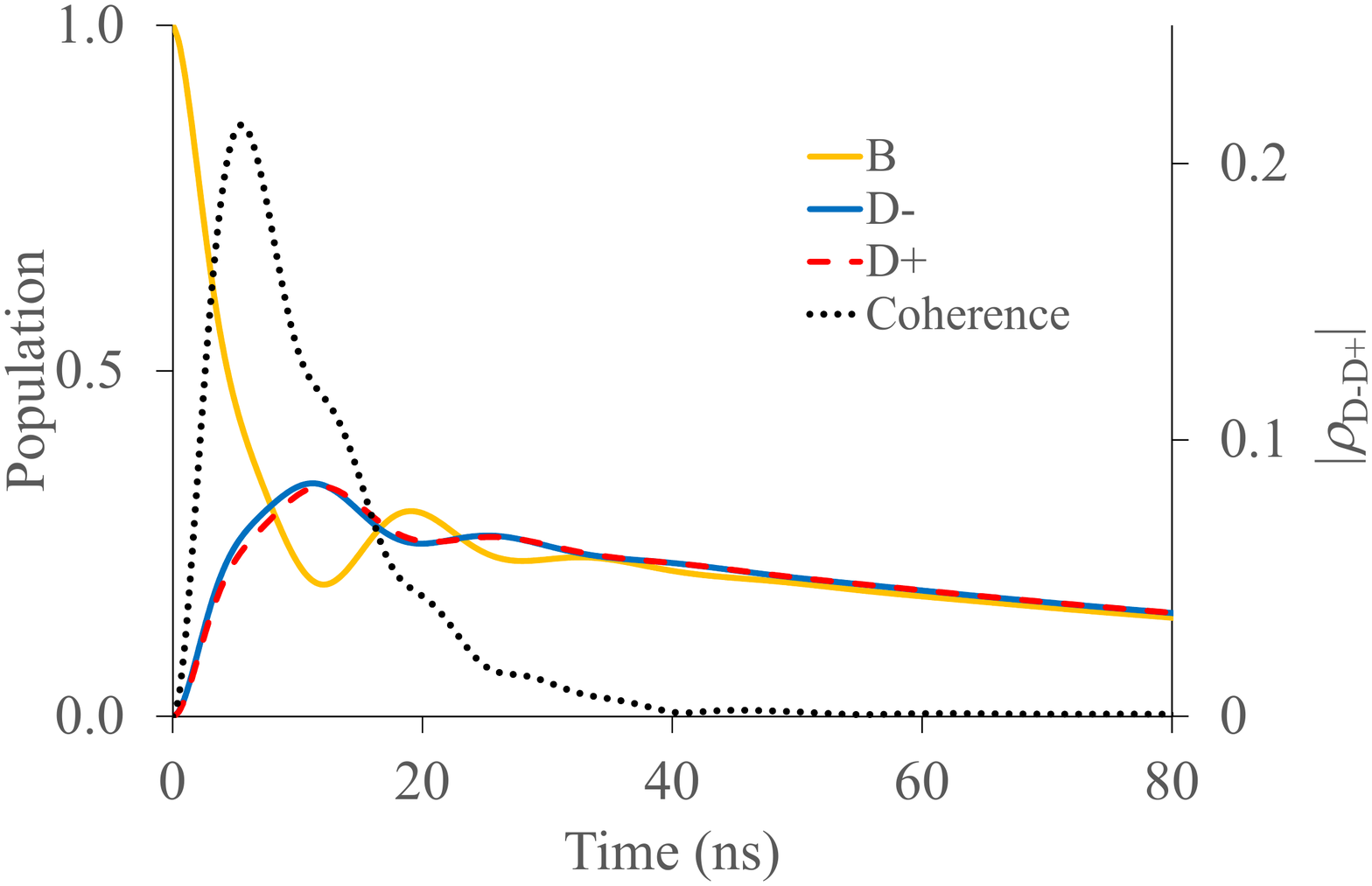}
\caption{(Color online)  Populations of the bright and lower dark doublet states (left vertical axis, light solid line for $B$, dark solid line for $D_-$, dotted line for $D_+$) and modulus of the coherence $\rho_{D_+D_-}(t)$ generated in the doublet (right vertical axis, dashed black line) as a function of time. Calculations are performed at room temperature, without field and without emission channel.}
\label{fig:popB}
\end{figure}

We first verify the calibration of the system-bath coupling in order to reproduce in the new energy domain the results of our previous work, i.e. the transitory generation of coherence in the dark lower doublet at room temperature $T$=298K from the bright state \cite{Chin_qubit2018}. Fig.\ref{fig:popB} shows the populations in the bright $B$ and dark doublet $D_+$ and $D_-$ states, together with the modulus of the coherence $\rho_{D_-D_+}(t)$ in this doublet (dotted line). As in our previous work, detailed balance is transiently broken due to the generation of coherence in the lower doublet, which can be seen in the population dynamics at $\sim 15$ns. The coherence lifetime is about 20ns and the slow decay of the populations due to spontaneous emission losses into the resonators may be seen at later times. 

\begin{figure}
  \centering
	\includegraphics[width =1.\columnwidth]{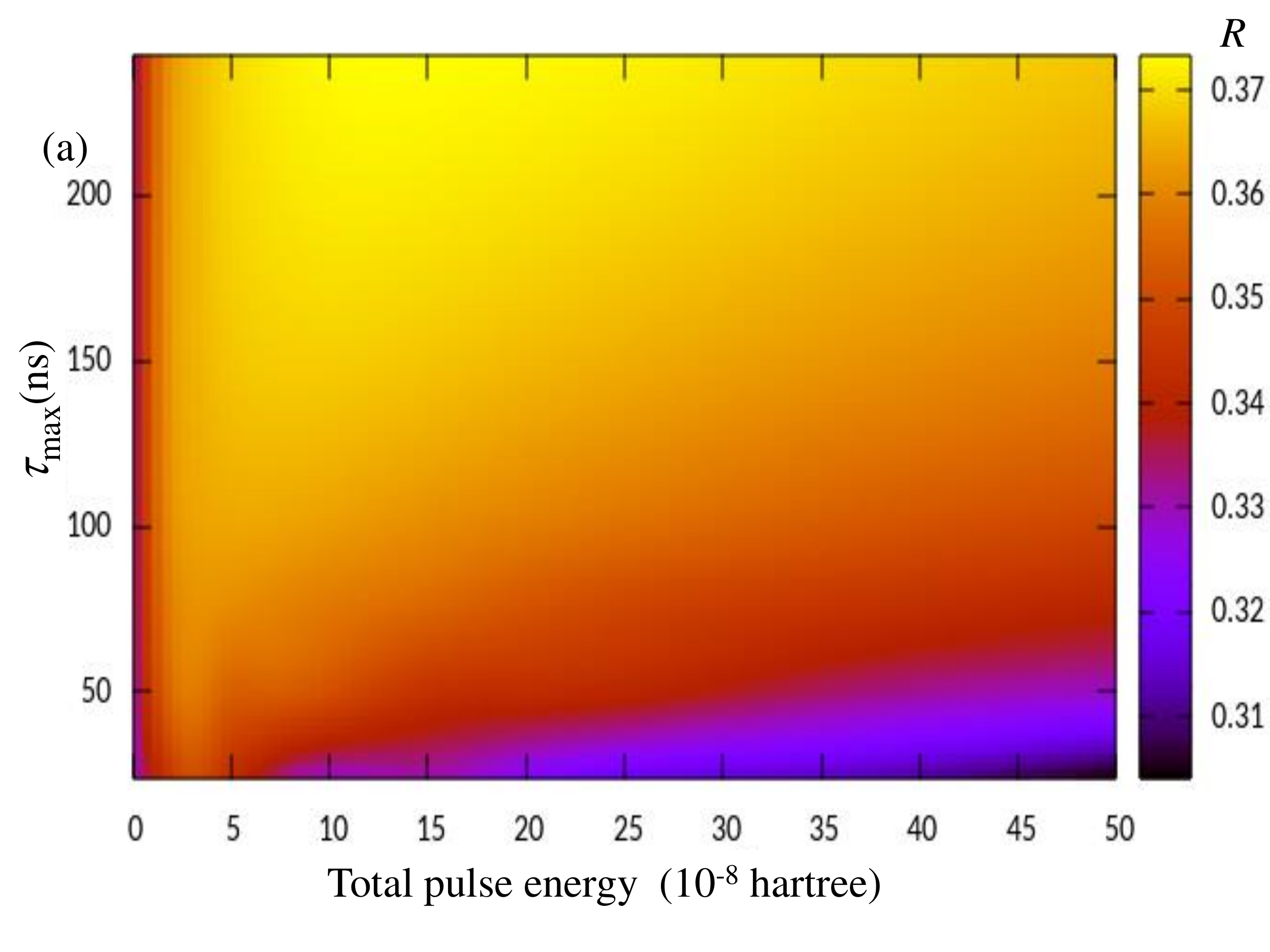}
	\includegraphics[width =1.\columnwidth]{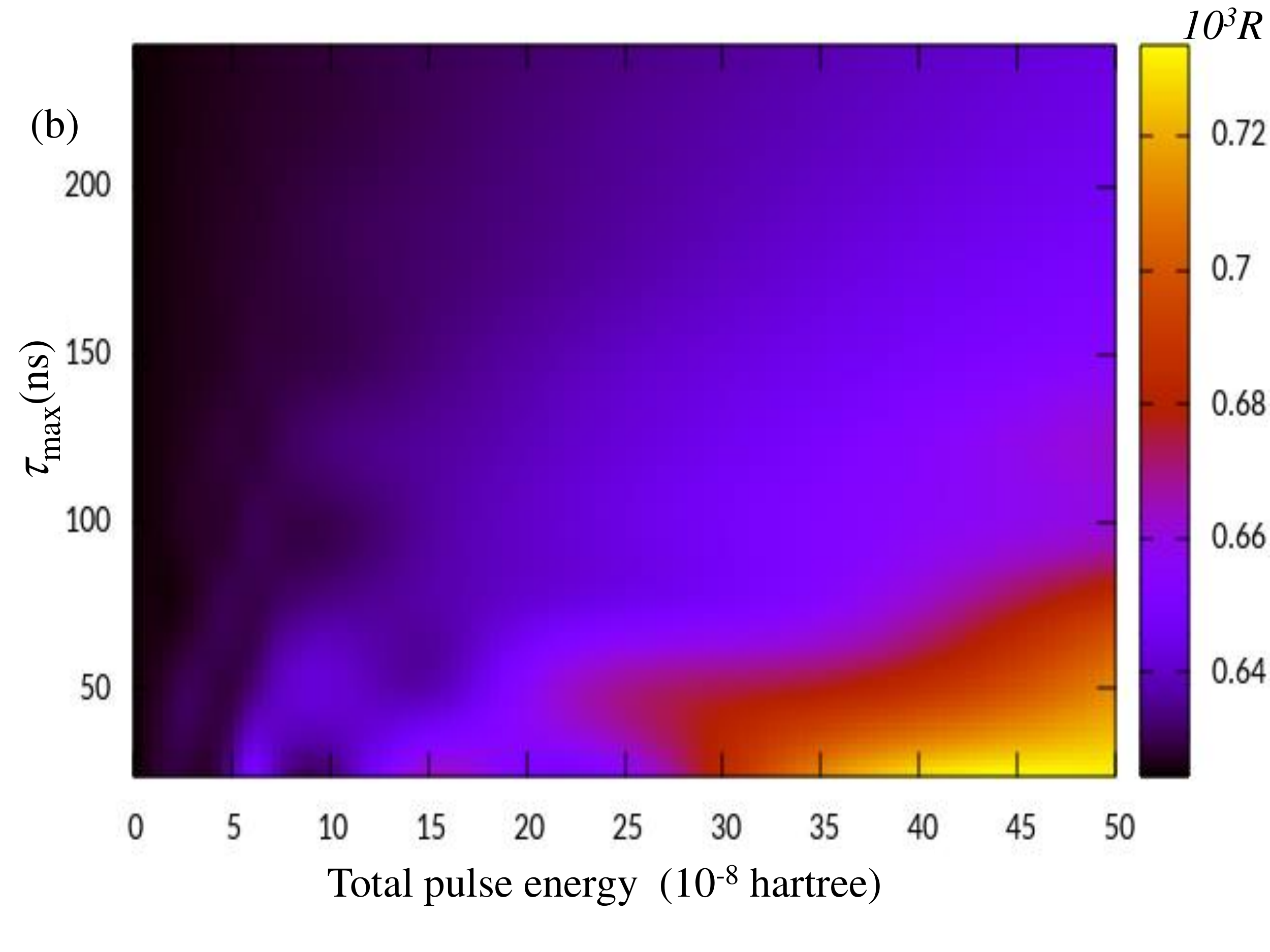}
\caption{ Iso-value contours of the ratio $R$  in Eq. (\ref{R}) as a function of the pulse energy Eq. (\ref{I}) and duration for classical noise at room temperature. Panel (a): $R$ taking into account the system-noise coupling, panel (b): $(10^3 R)$ without the bath.}
\label{fig:R298}
\end{figure}

In all the following simulations, the system is initially in its ground state and we take into account the spontaneous emission channel. Fig.\ref{fig:R298}a gives contour plots for the ratio $R$ as a function of the total energy $\mathcal{E}$ delivered by the pulse and its duration $\tau_{max}$. The range of variation of $R$ is rather small; namely, from 0.31 to 0.37. Highest values (i.e., best conditions to collect population in the resonator) are obtained for small pulse energies ($\mathcal{E}$ around $10^{-8}$Ha) associated with long pulse durations (around 200ns). We shall discuss this behavior below, by examining the dynamics induced by the different pulses. The lower panel of Fig.\ref{fig:R298}b shows $10^3 R$ when the ratio is obtained with same field characteristics, but without the system-noise coupling. The comparison of the two panels clearly shows the crucial role played by the bath when populating the excited state of ${{Q}_{3}}$ which finally decays into the resonator. In the absence of the bath $R$ is three orders of magnitude smaller: dissipation is thus essential for energy extraction in this system and we can expect efficiencies to be determined by noise properties.

\begin{figure}[!h]
\includegraphics[width =1.\columnwidth]{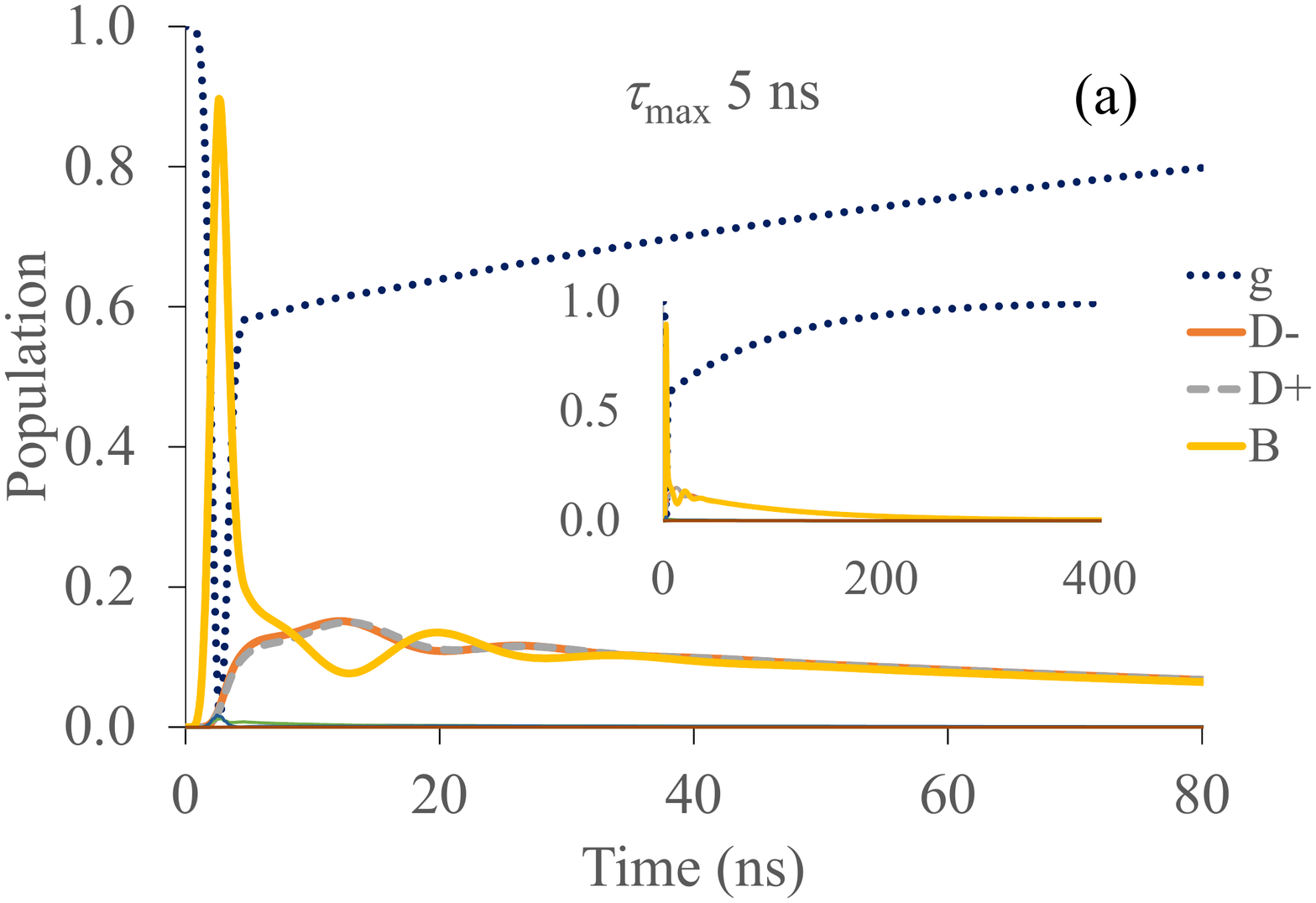}
\includegraphics[width =1.\columnwidth]{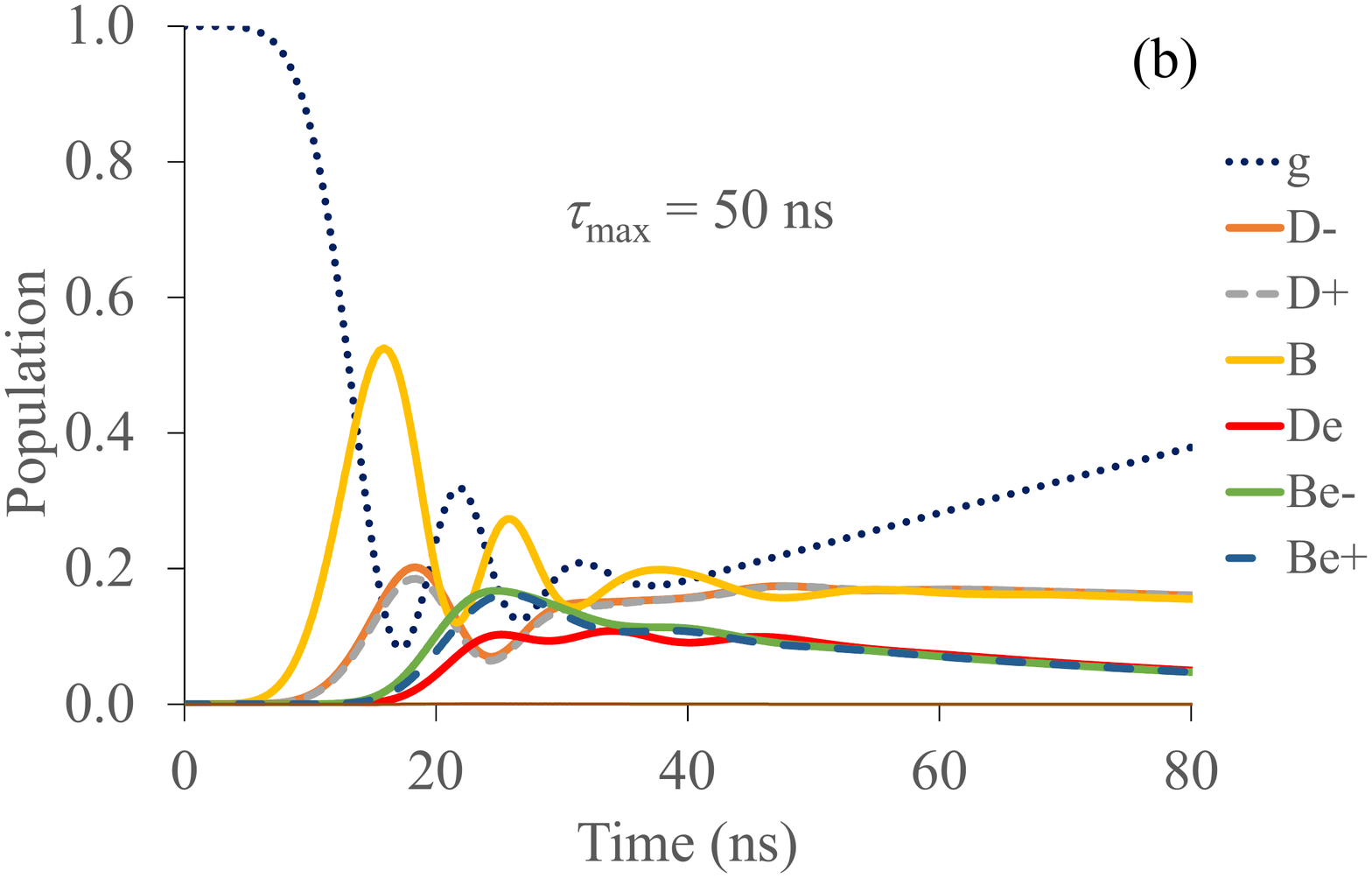}
\includegraphics[width =1.\columnwidth]{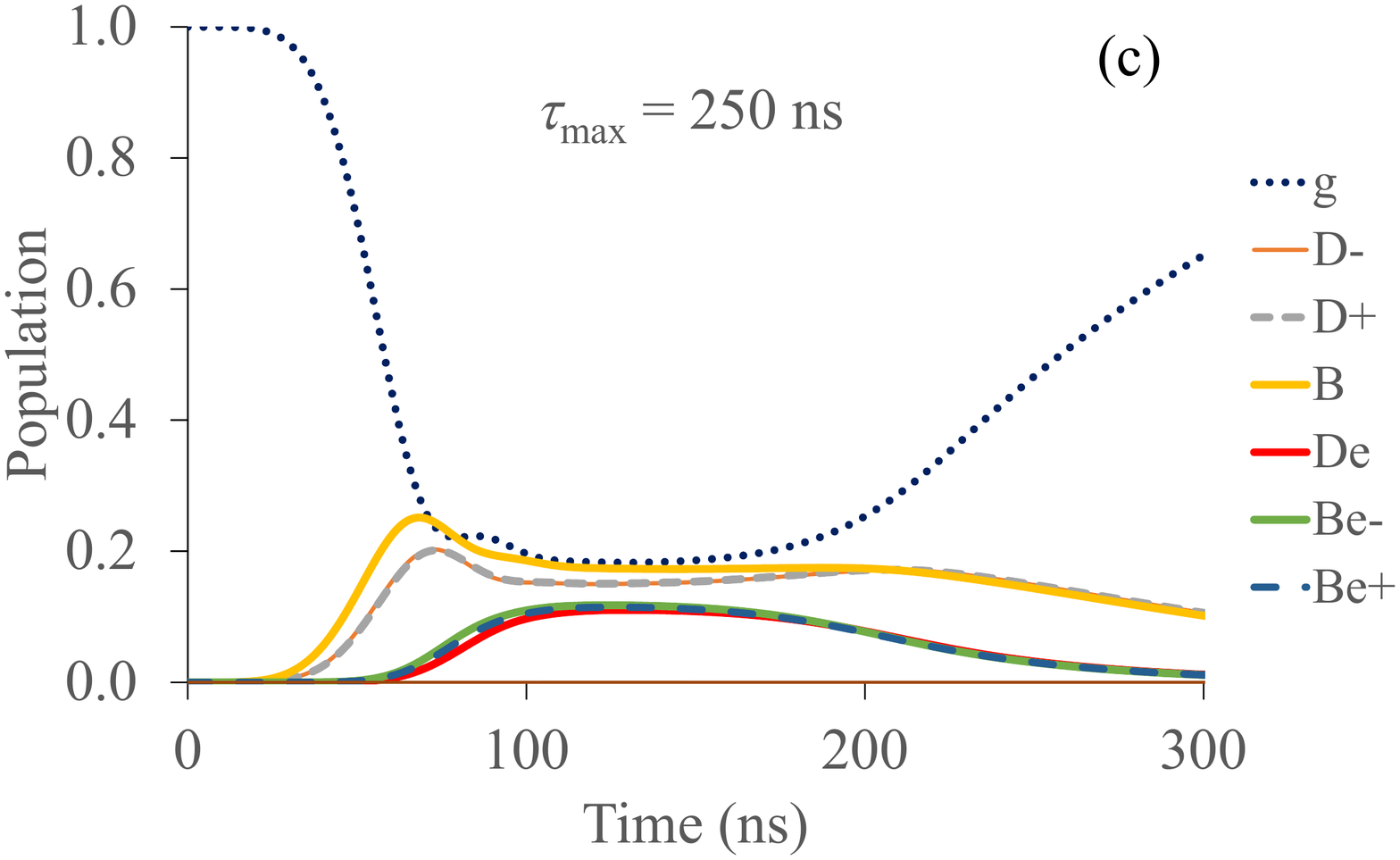}
\caption{Color online. Field driven dynamics at $T=$298$K$ (classical noise) for three pulses having the same integrated intensity $\mathcal{E}=5 \times 10^{-8}$Ha and different durations. }
\label{fig:dyn_I5}
\end{figure}

\begin{figure}
\includegraphics[width =1.\columnwidth]{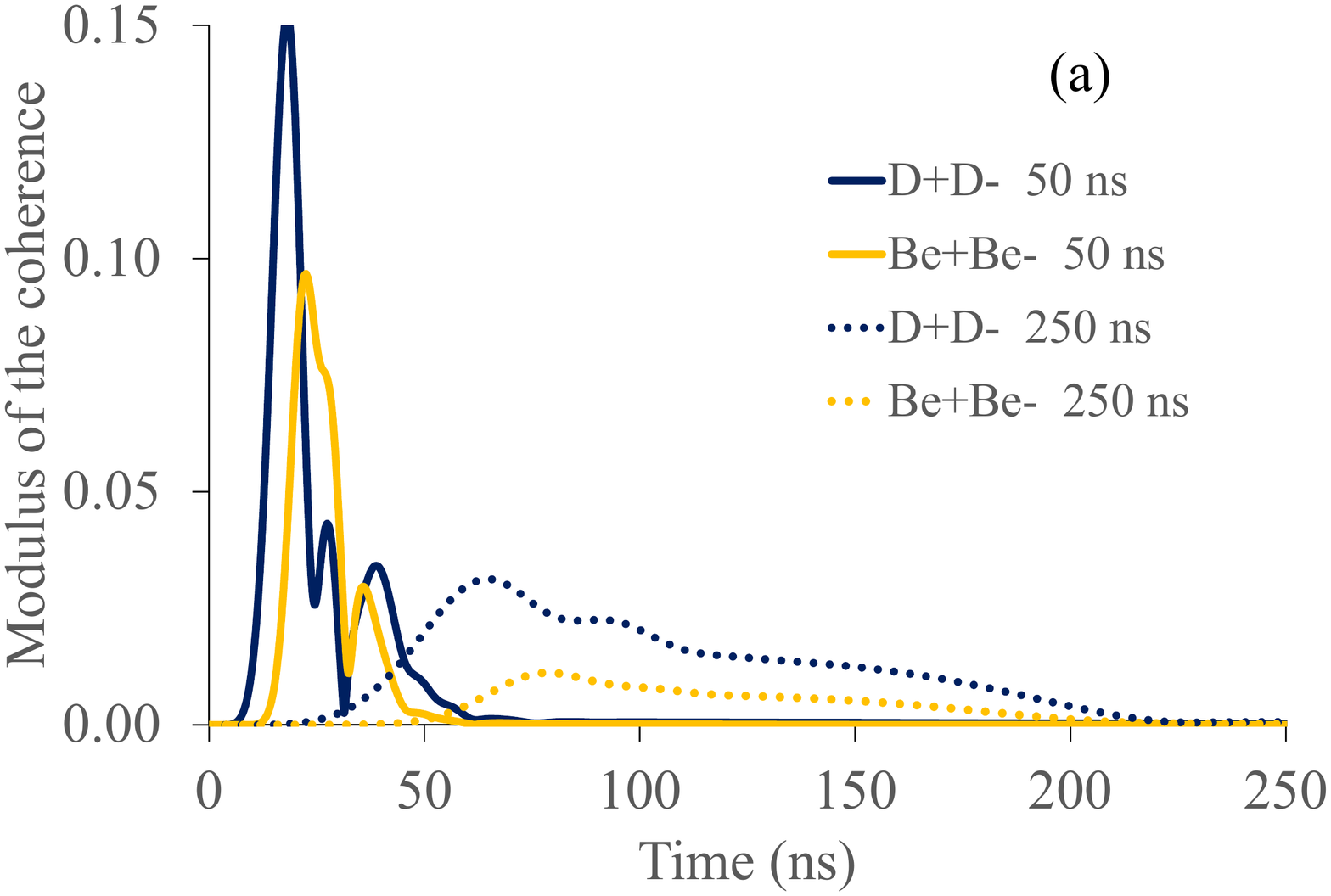}
\includegraphics[width =1.\columnwidth]{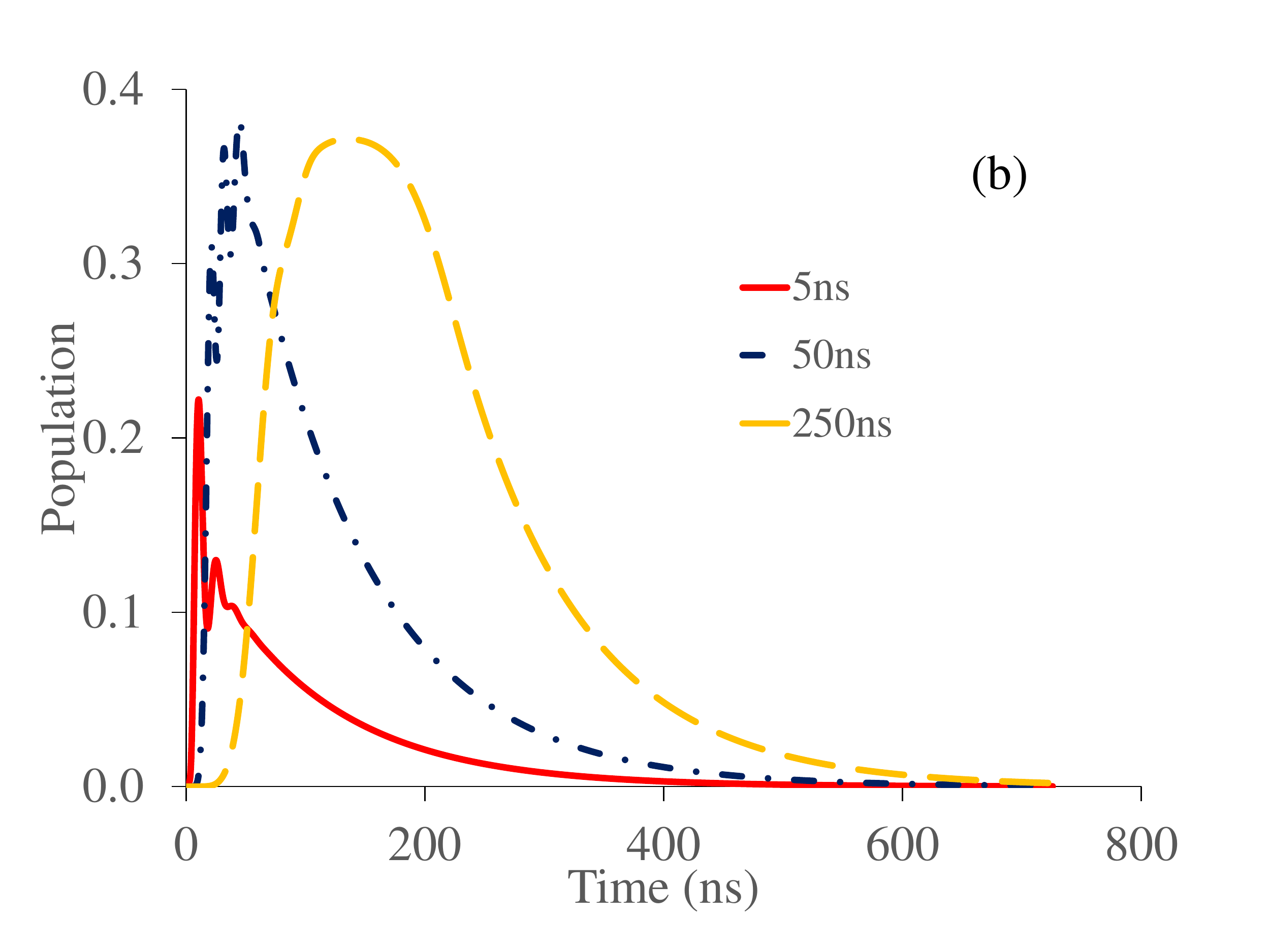}
\caption{ Color online. Panel (a) : Modulus of the coherence between the doublet states $D_{\pm}$ and $B_{e\pm}$ for two pulses with $\tau_{max}$=50 and 250ns and the same integrated intensity $\mathcal{E}=5 \times 10^{-8}$Ha at $T$=298K. Panel (b) : Population in the excited state of qubit $Q_3$ during the field driven dynamics for the same $\mathcal{E}$. }
\label{fig:coheP3_I5}
\end{figure}

In order to understand the variations of $R$, we first compare the effect of an increasing pulse duration for a weak pulse energy ( $\mathcal{E}=5 \times 10^{-8}$Ha). The population evolution is shown in Fig.\ref{fig:dyn_I5}. (i) The short pulse duration  $\tau_{max}$=5ns is smaller than the coherence decay time (about 20ns) of the field-free simulation (see Fig.\ref{fig:popB}). It basically acts as a $\pi$ pulse by populating the bright state $B$ which further decays by generating the lower dark doublet as in field-free case. The excited higher lying manifold is not populated. Full relaxation due to spontaneous emission occurs within 400ns (see Fig.\ref{fig:dyn_I5}a).  (ii) More interesting dynamics occur with a longer pulse of 50ns. One now observes several Rabi oscillations between $B$ and $g$ states and a transition towards the excited bright doublet ${{B}_{e\pm }}$ occurs from the lower dark doublet ${{D}_{\pm }}$ . This energy gap is also in resonance with the carrier frequency and the transition dipole moment is strong. There is a crossing between the populations of the dark ${{D}_{\pm }}$ and the bright doublets ${{B}_{e\pm }}$. On the other hand, the coupling via the bath induces a transition towards the dark excited state  ${{D}_{e}}$, which exhibits a slight oscillatory behavior revealing a weak back and forth transition with the excited doublet, before reaching the asymptotic mixture with equal weights in the three excited states. This is typical of a classical behavior with equal up and down transition rates. (iii) The very long pulse (250ns) induces yet another behaviour, with the occurrence of a steady state assisted by the field.  Rabi oscillations involving the bright $B$ and the ground $g$ states are completely damped. The three excited bright and dark states are populated simultaneously and a weak coherence in the doublets is sustained by the field. The modulus of the coherence in the doublets are shown in Fig.\ref{fig:coheP3_I5}a for the medium and long pulses. 

The coherence in the lower ${{D}_{\pm }}$ doublet (see Fig.\ref{fig:coheP3_I5}a) does not reach the maximum value of the field-free case since a transfer towards the excited doublet ${{B}_{e\pm }}$ takes place. The increase of the ${{B}_{e\pm }}$ coherence effectively rises when the one of ${{D}_{\pm }}$ decays. The switch is clearly seen in Fig.\ref{fig:coheP3_I5}a. The field assisted process is related with the steady state observed in the populations in Fig.\ref{fig:dyn_I5}. Note that the field driven coherence is several orders of magnitude larger than the one without the coupling to the bath which creates the early ${{D}_{\pm }}$ coherence. As discussed in our previous work \cite{Chin_qubit2018}, the main impact of this coherence generation is the population of the excited state of site $Q_3$ which transfers energy to the resonator. Fig.\ref{fig:coheP3_I5}b gives the population in the excited state of $Q_3$. As seen in the expression of the eigenvectors given in Appendix A, this state is populated both from the lower dark doublet ${{D}_{\pm }}$, from the upper dark state ${{D}_{e }}$ and from the excited doublet ${{B}_{e\pm }}$. This population increases with the pulse duration for a similar total energy delivered by the radiation. The losses due to spontaneous emission also increase leading to a rather small ratio $R$. 

We can now rationalize the qualitative evolution or the ratio $R$ with respect to the pulse duration at constant total energy $\mathcal{E}=5 \times 10^{-8}$Ha (weak increase, see Fig.\ref{fig:R298}a). $P_{res}$ increases with the population of the $Q_3$ excited state, but at the same time the excited doublet contributes to the spontaneous emission with the bright $B$ states so that $P_{loss}$ also increases. As a result, $R$  finally exhibits a weak variation with a slight domination of $P_{res}$.       

\begin{figure}
 \includegraphics[width =1.\columnwidth]{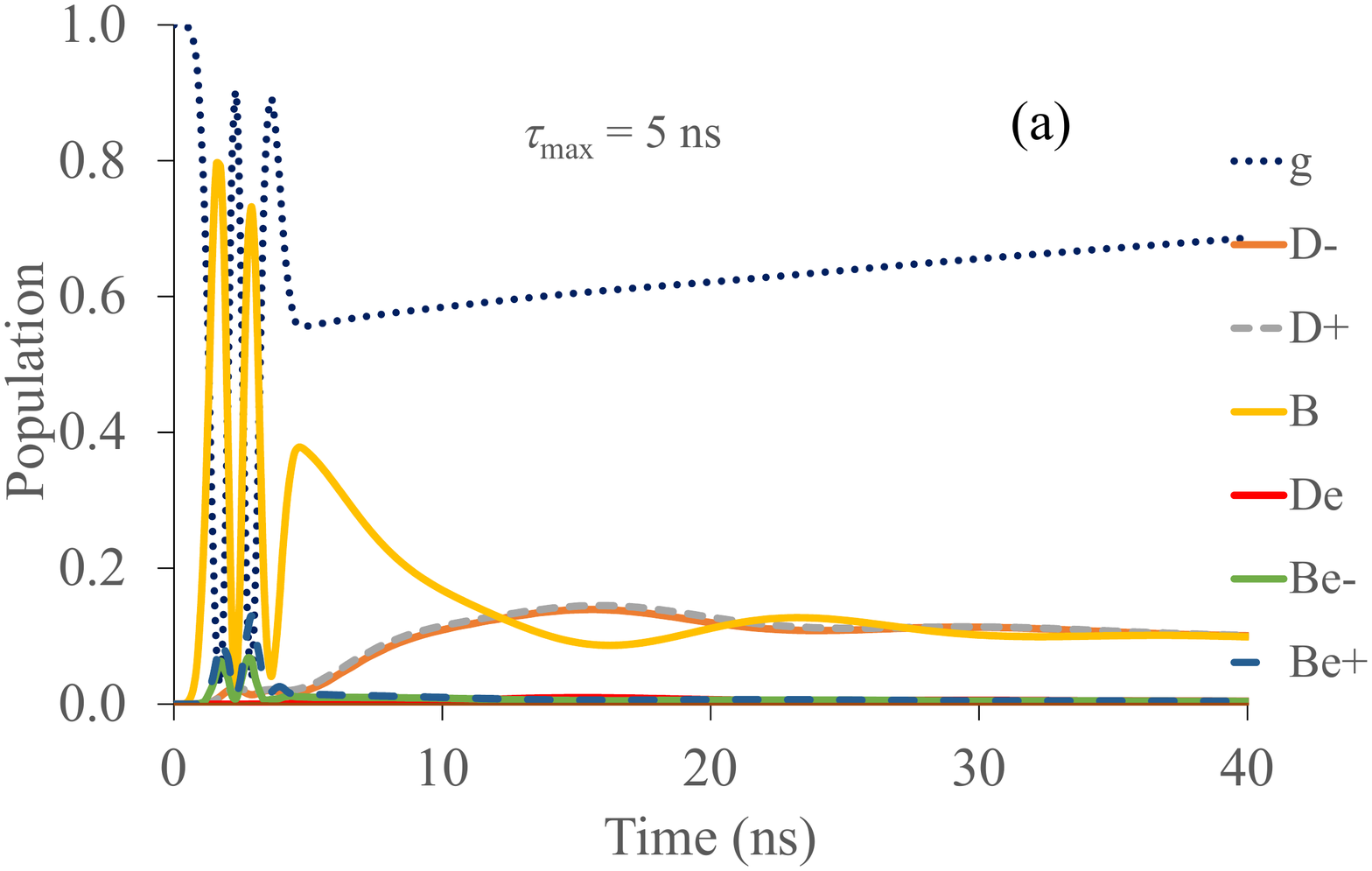}
 \includegraphics[width =1.\columnwidth]{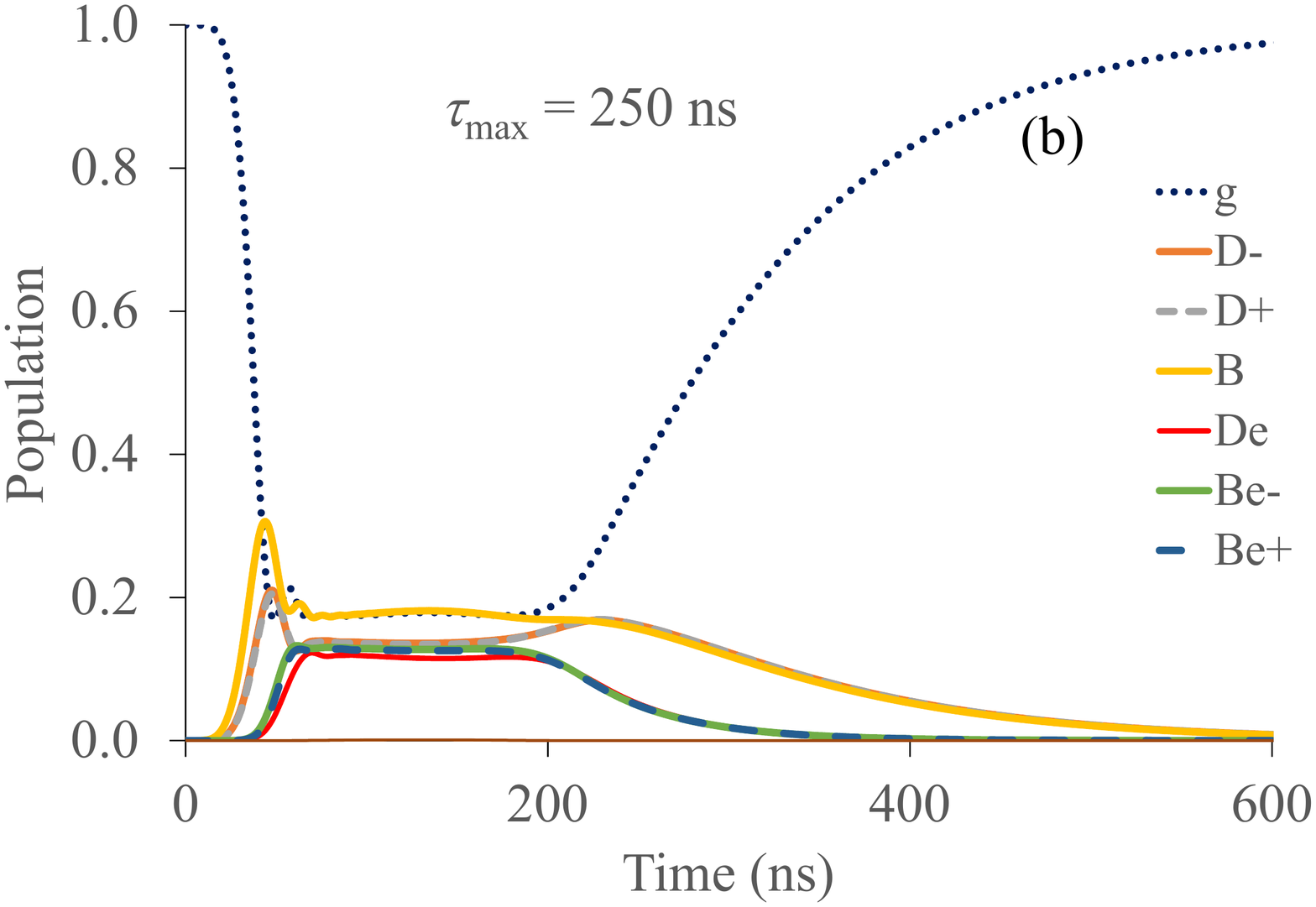}
\caption{ Color online. Field driven dynamics at $T=$298$K$ (classical noise) for two pulses having the same integrated intensity $\mathcal{E}=40 \times 10^{-8}$Ha and different durations}.
\label{fig:dym_I40}
\end{figure}

Fig.\ref{fig:dym_I40} illustrates the dynamics for a high pulse energy $\mathcal{E}=40 \times 10^{-8}$Ha and different pulse durations. (i) For a short pulse (5ns) the $R$ ratio decreases with the total energy of the pulse while for a long one (250ns) it remains more stable. By comparing Figs.\ref{fig:dyn_I5}a and \ref{fig:dym_I40}a for $\tau_{max}$=5ns, the main difference comes from the evolution of the brigth $B$ state which exhibits strong Rabi oscillations during the pulse for the high intensity while the populations in the lower doublet present the same profile and the upper excited manifold is not populated. The coherence generation in the lower dark doublet takes place during the field free dynamics so $P_{res}$ is stable while $P_{loss}$ increases due to the radiative decay of the $B$ state and finally $R$ decreases. (ii) With a very long pulse, as already observed at low intensity, the Rabi oscillations are early damped such that the population in the $B$ state remains low and also the radiative decay from this state. The field driven steady state with equal populations in the excited manifold leads to a similar behavior for any total intensity with the compensation between enhancement of population in $Q_3$ increasing $P_{res}$ and that of the bright doublet increasing $P_{loss}$ so that $R$ does not vary significantly.

\subsection{Quantum noise}
\label{qu_noise}
\begin{figure}
\centering
\includegraphics[width =1.\columnwidth]{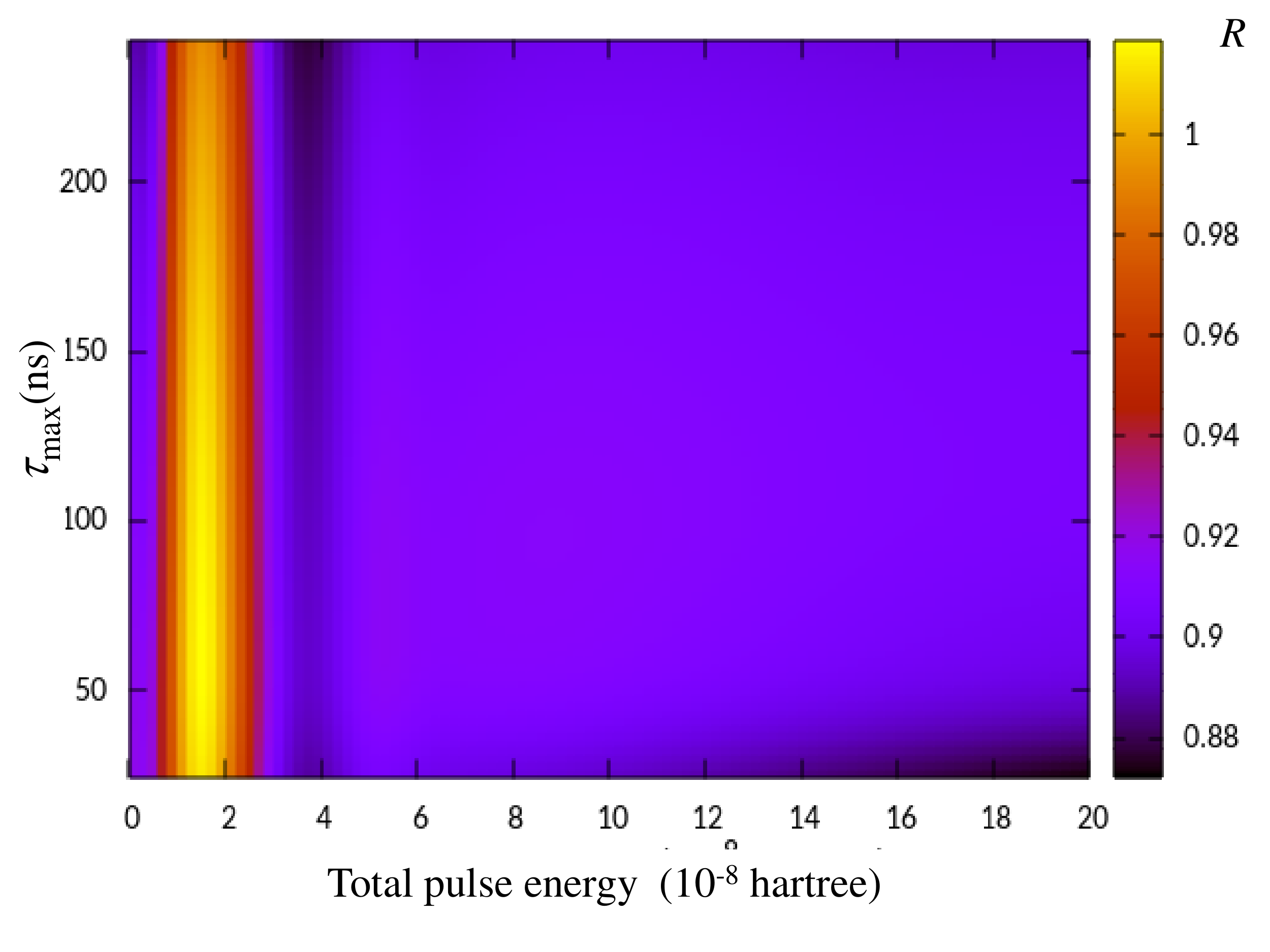} 
\caption{Color online. Iso-value contours for $R$ Eq.(\ref{R}) as a function of the pulse energy $\mathcal{E}$ Eq.(\ref{I}) and the pulse duration $\tau_{max}$ for quantum noise at $T$=0.01K.}
\label{fig:R001}
\end{figure}

Lowering the temperature down to $T$=0.01K leads to the quantum regime in the energy range under consideration. The correlation function Eq.(\ref{ct}) is complex and we recover the situation examined in ref. \cite{Chin_qubit2018} (taking into account the scaling factor $10^4$ between molecular and qubit regimes). We take again the coupling strength $\eta=0.01$ in Eq. (\ref{eta}).  Figure \ref{fig:R001} presents iso-value contours of the ratio $R$. In the quantum regime, $R$ is always close to 1, i.e., much higher than in the classical case. $P_{res}$ dominates $P_{los}$ which remains negligible. The highest ratio is for a low pulse energy ($\mathcal{E}=10^{-8}$Ha) and a duration smaller than 100ns. The key point to understand the difference of behavior is the rate of population or depopulation of the dark state $D_e$. In the classical regime, both rates up and down are equal so that even if the bath transfers population from the bright excited doublet $B_{e\pm}$ to the dark state $D_e$ the inverse process takes place leading to equal weights in all the three states. On the contrary, in the quantum regime, the uphill rate is lower than the downhill one and the population remains trapped in the $D_e$ dark state. This protects the system against the loss by emission of the excited doublet. 

\begin{figure}
\centering
 \includegraphics[width =1.\columnwidth]{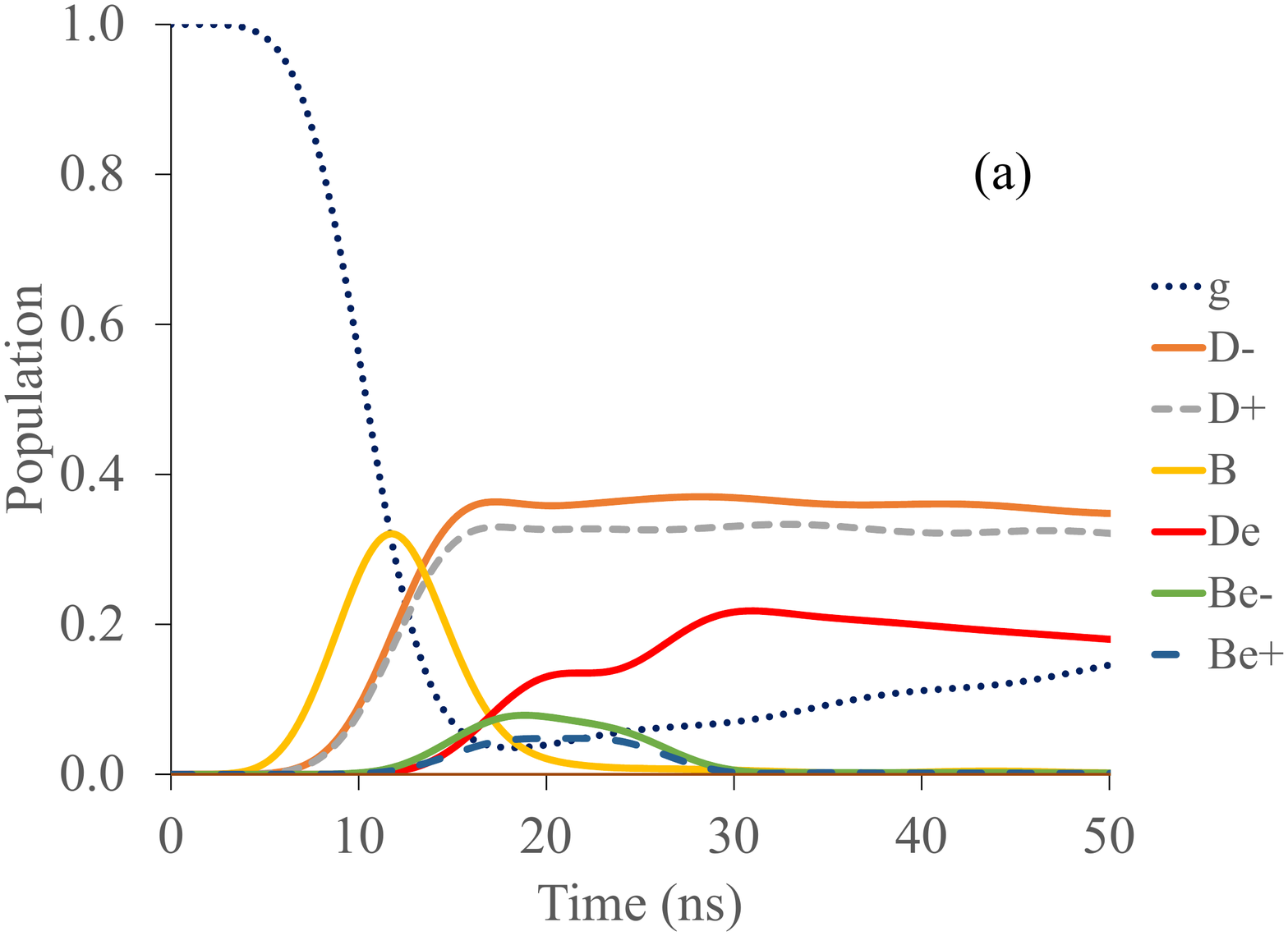}
\includegraphics[width =1.\columnwidth]{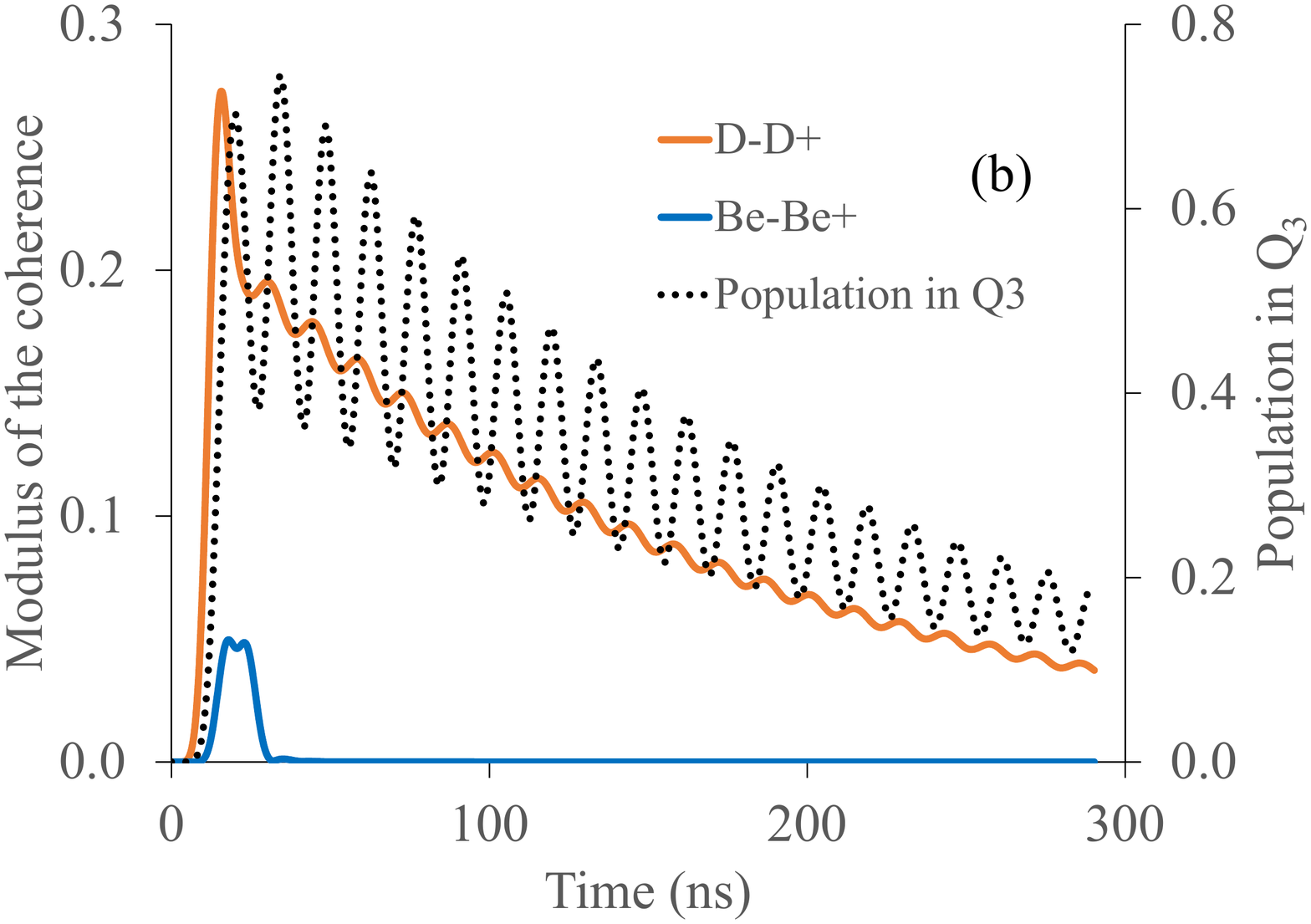}  
\caption{ Color online. Panel (a): Field driven dynamics at $T=$0.01$K$ for a pulse energy $\mathcal{E}=10^{-8}$Ha and $\tau_{max}$=25ns. The parameters $\eta$ is equal to 0.01. Panel (b), left axis : modulus of the coherence in the low and excited doublets; right axis : population in the excited state of $Q_3$. }
\label{fig:dyn_quant}
\end{figure}

This is illustrated in Fig.\ref{fig:dyn_quant} where dynamics are displayed for a pulse duration of 25ns and a low pulse energy $\mathcal{E}=10^{-8}$Ha In Fig.\ref{fig:dyn_quant}a, one observes the expected population of the lowest dark doublet generated by the bath and the subsequent excitation of the bright excited one but the latter rapidly decays via the bath towards $D_e$ without any re-population. The oscillating population in the excited state of $Q_3$ is shown in Fig.\ref{fig:dyn_quant}b (right axis). It follows the coherence in the dark doublet. For a higher energy, the $R$ ratio slightly decreases towards about 0.9. This is due to the fact that a high intensity induces Rabi oscillations which enhances the loss from the bright state only. This is similar to the process already obtained in the classical behavior. 

\section{Discussion and Conclusions}

By introducing an explicit treatment of the light-matter interactions in the form of MW wave guide driving and spontaneous radiative losses, we have elucidated a number of useful mechanisms for small-molecule light harvesting units. For classical noise, as is presently available in qubit simulators, we have shown that resonant driving of the main ground state optical transition also leads to strong excitation of the doubly-excited manifold. However, as classical (`infinite temperature') noise rapidly equalises the populations of all the states within each excitation number sector, the efficiency quantified by the parameter $R$ in Eq.(\ref{R}) tends to decrease with pulse length due to the increased emission of the doubly excited states \emph{back} into the wave guide. Under classical noise conditions we thus find that the efficiency of this three-qubit energy harvesting system is effectively set by the ratios of the spontaneous emission rates into the wave guide and resonator ($\sim 0.3$, as observed in the experiments of Ref. \cite{potovcnik2018studying}). However, although the efficiency does not increase, we should point out that the inclusion of higher excited states \emph{does} lead to a larger absolute value of energy capture and transfer as a function of power and pulse length. We also note that the spontaneous generation of coherent quantum motion from noise, as predicted in Ref. \cite{Chin_qubit2018}, is again observed in \emph{both} the single and double excitation sectors of our expanded and more realistic model, and thus should be observable in time-resolved experiments on the setup of Ref. \cite{potovcnik2018studying}.  The noise in the simulator of Poto{\v{c}}nik $\it{et}$ $\it{al.}$ is classical and can take spectral form that can be produced by commercially available signal generators. It is therefore possible that other types of  EET models  might be explored, such as those proposed by Briggs and Eisfeld \cite{eisfeld2011,eisfeld2012} with classical real Markovian noise. It should be also very interesting to check the present simulation with driving field and non-Markovian noise with the classical master equation these authors have proposed since this classical strategy is able to describe coupling of populations and coherences in the eigenbasis which is at the hard of the process studied here \cite{Chin_qubit2018}.

When the qubits are subject to a low temperature quantum environment, we find that rapid and effectively unidirectional relaxation suppresses the population of all bright states in both the single and double excitation sectors. Indeed, the energy levels, eigenfunctions and allowed optical transitions of this quantum circuit lead to a ratchet-like situation, as depicted in Fig.\ref{fig:dyn_quant}b. This structure allows the lowest dark states to absorb photons from the wave guide, with dissipation rapidly `storing' this extra quantum in a doubly excited state that cannot emit back into the wave guide. Such 'quantum noise' is presently not available in superconducting simulators of light-harvesting processes, but could be realised with linear circuit components that simulate quantum harmonic oscillators \cite{Mostame2012}. Moreover, it would be interesting to see if this effective eigenstate structure could be extended to larger numbers of qubits and larger throughputs of photons while still maintaining $R\sim 1$.\cite{zhang2015delocalized} In this context, we also believe that it would also be insightful to consider the full counting statistics of photon emission into the resonator under classical and quantum noise, and shall take up this task in a forthcoming work.

\appendix

\section{Methods of dissipative quantum dynamics}
\label{appendix:METH}
The HEOM formalism is currently well documented and we summarize only the operational equations for the particular parametrization of the spectral density adopted here. With the expressions (\ref{Corre}) and (\ref{Correconj}), the master equation is written as a time-local hierarchical system of coupled differential equations among auxiliary operators:
\begin{align}
  & \overset{\bullet }{\mathop{{{\rho }_{\mathbf{n}}}}}\,\left( t \right)=-i\left[ {{H}_{S}}+{{H}_{f}}(t)+{{H}_{ren}},{{\rho }_{\mathbf{n}}}(t) \right]+i\sum\limits_{k=1}^{{{n}_{cor}}}{{{n}_{k}}{{\gamma }_{k}}{{\rho }_{\mathbf{n}}}(t)} \nonumber \\ 
 & -i\left[ S,\sum\limits_{k=1}^{{{n}_{cor}}}{{{\rho }_{\mathbf{n}_{k}^{+}}}(t)} \right]-i\sum\limits_{k=1}^{{{n}_{cor}}}{{{n}_{k}}\left( {{\alpha }_{k}}S{{\rho }_{\mathbf{n}_{k}^{-}}}-{{{\tilde{\alpha }}}_{k}}{{\rho }_{\mathbf{n}_{k}^{-}}}S \right)} \nonumber \\ 
 & +{{L}_{3}}\left[ {{\rho }_{\mathbf{n}}}(t) \right]{{\delta }_{\mathbf{n},1}}+{{L}_{12}}\left[ {{\rho }_{\mathbf{n}}}(t) \right]{{\delta }_{\mathbf{n},1}}.  
\end{align}	
The auxiliary operators are denoted by a collective index $\mathbf{n}=\left\{ {{n}_{1}},\cdots ,{{n}_{{{n}_{cor}}}} \right\}$ where ${{n}_{j}}$ is the quantum number giving the excitation in the pseudo mode $j$ ($j=1,{{n}_{cor}}$) of the correlation function. The system density matrix is given by the first row, i.e. $\mathbf{n}=\left\{ 0,\cdots ,0 \right\}$ hence ${{\rho }_{S}}(t)={{\rho }_{1}}\left( t \right)={{\rho }_{\left\{ 0,\cdots ,0 \right\}}}\left( t \right)$. The level of the hierarchy is equal to the sum of the quantum numbers of the modes. $\mathbf{n}_{k}^{\pm }=\left\{ {{n}_{1}},\cdots ,{{n}_{k}}\pm 1,\ldots ,{{n}_{{{n}_{cor}}}} \right\}$ is the index of the auxiliary operator for which the pseudo mode $k$ has been excited or de-excited by one quantum (each matrix is coupled only with the next superior and inferior level in the hierarchy).

The Lindblad operators, involving emission rates $G_k$, are built from the raising ${{d}_{+(k)}}$ and lowering ${{d}_{-(k)}}$ operators, where $k=3$ represents the ${{Q}_{3}}$ qubit and $k=12$ corresponds to the sum of the ${{Q}_{1}}+{{Q}_{2}}$ qubits, i.e., ${{d}_{\pm (12)}}={{d}_{\pm 1}}+{{d}_{\pm 2}}$
\begin{align}
  & {{L}_{k}}\left[ \rho (t) \right]=\frac{{{G}_{k}}}{2}\left[ 2{{d}_{-(k)}}\rho (t){{d}_{+(k)}} \right. \nonumber \\ 
 & \left. -{{d}_{+(k)}}{{d}_{-(k)}}\rho (t)-\rho (t){{d}_{+(k)}}{{d}_{-(k)}} \right]  
\end{align}
The HEOM equations are solved in the interaction representation by the Cash-Karp Runge-Kutta algorithm with adaptative time step \cite{numrec_1992}. The initial time step is $10^5$a.u. It is well known that the HEOM formalism may become inaccurate at low temperature because a very large number of Matsubara terms are then required \cite{cao2013}. However, in the present application, the thermal energy at $T = 0.01K$ is still of the order of magnitude of the main energy gaps so that this chosen temperature is not so small and in practice. The number of Matsubara terms remains small for the quantum simulation at low temperature. Convergence has been checked by increasing the number of Matsubara terms in some examples. 5 Matsubar terms are largely enough. To ensure convergence in any case, we used level 5 for the HEOM hierarchy and 10 Matsura terms.  In the classical regime at room temperature, the HEOM level is four and the number of Matsubara is 1 but 0 should be enough.

\section{Eigenvectors of lower and upper triplets}
\label{appendix:EV}

The eigenstates of ${{H}_{S}}$ form two important triplets, the ground cluster contains a bright state $\left| B \right\rangle $ which can be excited by the transmission line and a doublet of dark states $\left| {{D}_{\pm }} \right\rangle $. The corresponding eigenvectors are given by:
\begin{eqnarray}
\left| B \right\rangle =0.71\left| 010 \right\rangle +0.71\left| 100 \right\rangle -0.035\left| 001 \right\rangle  \nonumber \\ 
\left| {{D}_{+}} \right\rangle =0.50\left| 010 \right\rangle -0.50\left| 100 \right\rangle +0.7\left| 001 \right\rangle \nonumber \\
\left| {{D}_{-}} \right\rangle =0.50\left| 010 \right\rangle -0.50\left| 100 \right\rangle -0.71\left| 001 \right\rangle 
\end{eqnarray}
In the excited triplet, the characteristics of the states are inverted leading to a bright doublet and a dark isolated state. The excited eigenvectors are:
\begin{eqnarray}
\left| {{D}_{e}} \right\rangle =-0.71\left| 011 \right\rangle +0.71\left| 101 \right\rangle -0.035\left| 110 \right\rangle \nonumber \\ 
\left| {{B}_{e-}} \right\rangle =-0.50\left| 011 \right\rangle -0.50\left| 101 \right\rangle +0.70\left| 110 \right\rangle  \nonumber \\
\left| {{B}_{e+}} \right\rangle =-0.50\left| 011 \right\rangle -0.50\left| 101 \right\rangle -0.71\left| 110 \right\rangle  
\end{eqnarray}

\section{Spectral density parameters}
\label{appendix:spectraldensity}
The parameters of the superohmic expression (Eq.(\ref{J})) for the spectral density used in the HEOM simulations are gathered in the following table. The $p$ parameter is $2.7959\times10^{-47}$a.u. in the classical case ($T = 298K$) and $2.7959\times10^{-41}$a.u. in the quantum case $T = 0.01K$.  

\begin{table}[ht]
 {\begin{tabular}{cccc} \toprule
		 $\Omega_{1}$ (a.u.)    & $\Gamma_1$ (a.u.)   & $\Omega_2$ (a.u.) & $\Gamma_2$ (a.u.) \\[0.5ex]
	\hline \noalign{\vskip 1.0ex}
		$3.1892\times10^{-8} $  & $2.1191\times10^{-7} $  & $1.5222\times10^{-7} $  & $9.0678\times10^{-9}$ \\[0.5ex]
	\end{tabular}}
\label{table:SD_values}
\end{table}

\section*{Acknowledgment}
		
Alex Chin acknowledges for the Jean d'Alembert Chaire from IDEX Paris-Saclay, contrat CNRS 157819. Etienne Mangaud acknowledges support from the ANR-DFG COQS, under Grant No. \mbox{ANR-15-CE30-0023-01.} This work has been performed within the French GDR 3575 THEMS.

\end{document}